\pdfoutput=1 

\documentclass[
     a4paper,
     ]{article}

\usepackage{IEK10} 
\usepackage{natbib}
\usepackage{afterpage}
\DeclareUnicodeCharacter{03B2}{\ensuremath{\beta}}
\DeclareUnicodeCharacter{03B1}{\ensuremath{\alpha}}
\DeclareUnicodeCharacter{03C9}{\ensuremath{\omega}}
\DeclareUnicodeCharacter{2212}{-}

\newcommand{\mytitle}{Predicting the Temperature-Dependent CMC of Surfactant Mixtures with Graph Neural Networks}

\newcommand{\affil}{
  \begin{itemize}[leftmargin=3mm, itemsep=0mm]
        \item[$^a$]BASF Personal Care and Nutrition GmbH, Henkelstrasse 67, 40589 Duesseldorf, Germany %
        \item[$^b$]RWTH Aachen University, Process Systems Engineering (AVT.SVT), 52074 Aachen, Germany %
		\item[$^c$]Forschungszentrum J\"ulich GmbH, Institute for Energy and Climate Research IEK-10: Energy Systems Engineering, 52425 J\"ulich, Germany%
		\item[$^d$]JARA Center for Simulation and Data Science (CSD), 52056 Aachen, Germany
  \end{itemize}
}

\def\firstAuthor{Christoforos Brozos}
\newcommand{\myauthor}{
	Christoforos Brozos$^{a,b}$, 
	Jan G. Rittig$^b$,
    Elie Akanny$^a$,
	Sandip Bhattacharya$^a$,
    Christina Kohlmann$^a$,
	Alexander Mitsos$^{d,b,c,*}$ %
}
\newcommand{\correspondingauthor}{
    Corresponding author, E-mail: amitsos@alum.mit.edu
}
\author{\myauthor}

\usepackage[hyphens]{url}
\usepackage[
  colorlinks,
  linkcolor=blue,
  citecolor=blue,
  urlcolor=blue,
  pdftitle={\mytitle},
  pdfauthor={\firstAuthor}
]{hyperref}
\usepackage[capitalise, nameinlink]{cleveref}
\crefname{table}{Tab.}{Tab.}

\begin{document}

	\thispagestyle{firststyle}
	
	\begin{center}
		\begin{large}
			\textbf{\mytitle}
		\end{large} \\
		\vspace{0.1cm}
		\myauthor
	\end{center}
	
	\vspace{-0.4cm}
	
	\begin{footnotesize}
		\affil
	\end{footnotesize}
	
	\vspace{-0.3cm}

	\section*{Abstract}
	
	Surfactants are key ingredients in foaming and cleansing products across various industries such as personal and home care, industrial cleaning, and more, with the critical micelle concentration (CMC) being of major interest. Predictive models for CMC of pure surfactants have been developed based on recent ML methods, however, in practice surfactant mixtures are typically used due to to performance, environmental, and cost reasons. This requires accounting for synergistic/antagonistic interactions between surfactants; however, predictive ML models for a wide spectrum of mixtures are missing so far. Herein, we develop a graph neural network (GNN) framework for surfactant mixtures to predict the temperature-dependent CMC. We collect data for 108 surfactant binary mixtures, to which we add data for pure species from our previous work [Brozos et al. (2024), J. Chem. Theory Comput.]. We then develop and train GNNs and evaluate their accuracy across different prediction test scenarios for binary mixtures relevant to practical applications. The final GNN models demonstrate very high predictive performance when interpolating between different mixture compositions and for new binary mixtures with known species. Extrapolation to binary surfactant mixtures where either one or both surfactant species are not seen before, yields accurate results for the majority of surfactant systems. We further find superior accuracy of the GNN over a semi-empirical model based on activity coefficients, which has been widely used to date. We then explore if GNN models trained solely on binary mixture and pure species data can also accurately predict the CMCs of ternary mixtures. Finally, we experimentally measure the CMC of 4 commercial surfactants that contain up to four species and industrial relevant mixtures and find a very good agreement between measured and predicted CMC values.
	
	\vspace{0.1cm}



\section{Introduction}\label{sec:introduction}

\noindent Surfactant mixtures commonly exhibit advantageous synergistic properties compared to single surfactants and therefore are preferred in many applications~\citep{Rosen2012,Rosen1982,Huang2017,KumarShah2022,Geng2017,Moulik2021,Cheng2020}, such as personal care products, detergents, pharmaceuticals and others~\citep{Shaban2020,Nitschke2007,Massarweh2020,2005,Hunter1998,RodriguezPatino2008}. Commercial formulations developed in cosmetics and detergents industries contain almost exclusively surfactant mixtures~\citep{MyersAugust2020,Grady2023,Kelleppan2023}. This is either due to the existence of a homologous distribution in an industrial grade surfactant or the combination of surfactants driven by performance, cost, and sustainability aspects. Therefore, understanding and modeling surfactant mixtures is essential for the design of further tailored products.

\par The properties of surfactant mixtures highly depend on the interaction effects between the surfactant structures. Surfactants are amphiphilic molecules composed of a hydrophilic (head) and a hydrophobic (tail) part. An important property of a surfactant mixture is the critical micelle concentration (CMC), which denotes the minimum surfactant concentration causing an emergence of surfactant micelles. 
Surfactant mixtures are generally described as complex systems due to the difference between bulk and micelle concentration of the pure species, i.e., mixture components~\citep{Grady2023,Clint1975}. Mixing two surfactants may lead to a synergistic or antagonistic behavior. A binary mixture exhibits synergism if at any molar fraction the mixture CMC, denoted as $CMC_\text{mix}$, is lower than the CMC of either pure species, and antagonism if $CMC_\text{mix}$ is higher than the CMC of either pure species~\citep{Rosen2012,Alargova2001}. A smaller CMC indicates that a lower surfactant concentration is needed to form micelles and therefore desired for cleansing applications. 

\par The mixing behavior of surfactants is influenced by factors such as the surfactant molecular structure, temperature, presence of electrolytes and the pH. 
Herein, we consider the influence of the surfactant molecular structure and temperature.
Binary mixtures between two nonionic surfactants have shown to form ideal micelles, i.e., the two pure species are mixed ideally on the micelle~\citep{Holland1983}, which is due to the lack of electrostatic repulsive forces between the head groups~\citep{Huang2017,MyersAugust2020,Zhang2004}. Nonionic/anionic and nonionic/cationic surfactant systems tend to mix nonideally, i.e., nonideal mixing in the micelle, and to behave synergistically~\citep{Rosen1982,KumarShah2022,Grady2023,Zhang2004,Ren2014,Haque1996,Hines1997,JeongSoo2004}. However, not all mixtures between nonionic and ionic surfactants exhibit synergism, as was shown by Alargova et al.~\citep{Alargova2001}. The formation of micelles in anionic/cationic mixtures benefits from the reduction of the repulsive forces between the head groups and thus a synergistic behavior is observed~\citep{Grady2023,Phaodee2021}. Anionic/anionic mixtures are shown to exhibit an antagonistic micellar behavior~\citep{L.LopezFontan1999}, due to large repulsive forces between the head groups in surfactant micelles. Zwitterionic surfactants have a different charge depending on the pH of the solution, and hence mixtures containing them can exhibit multiple behaviors based on the solution conditions. Attractive interactions between anionic and zwitterionic surfactants have been reported in the literature~\citep{Hines1998,Hines1997a,McLachlan2006}; similarly, slight deviation from ideal mixed micelles between cationic and zwitterionic surfactants is found~\citep{McLachlan2006}. 
For more details on possible interactions between surfactant classes, we refer to reviews in Refs.~\citep{KumarShah2022,Grady2023,Phaodee2021,Abe2004}.
Overall, binary surfactant mixtures are complex systems and their behavior is not fully understood yet.

\par The importance of the $CMC_\text{mix}$ for many applications has motivated the development of mathematical models throughout the years. For systems where micelle formation is ideal, such as nonionic/nonionic combinations, John H. Clint~\citep{Clint1975} proposed the following equation to calculate the $CMC_\text{mix}$ of a binary mixture based on the mole fractions, $x_{1}$ and $x_{2}$, and the CMCs of the two surfactants as pure species: 
\begin{equation}
    \frac{1}{CMC_\text{mix}} =  \frac{x_{1}}{CMC_\text{1}} + \frac{x_{2}}{CMC_\text{2}}
    \label{eqn:clint_method}
\end{equation}

\noindent To account for nonideal micelle formation, the activity coefficients ($\gamma_{1}$, $\gamma_{2}$) of surfactants 1 and 2 are introduced, leading to the following equation:
\begin{equation}
    \frac{1}{CMC_\text{mix}} =  \frac{x_{1}}{\gamma_{1}~CMC_\text{1}} + \frac{x_{2}}{\gamma_{2}~CMC_\text{2}},
    \label{eqn:activity_coef}
\end{equation}

\noindent Note the inverse relationship on the activity coefficients in contrast to the usual  multiplication of activity coefficients and mole fractions. One option to calculate the activity coefficients was proposed by Rubingh based on the regular solution theory~\citep{Holland1983,Rubingh1979}. Rubingh's model introduces the interaction parameter $\beta$ as a way to account for the nonideal mixed micelle formation, which can be estimated from the experimentally obtained CMC values of the pure species and at least one mixture of them. Knowing $\beta$, allows us to calculate the activity coefficients by~\citep{Holland1983}: 
\begin{equation}
    \gamma_{1} = \exp(\beta \cdot x_{2}^{2})
    \label{eqn:gama_1}
\end{equation}

\begin{equation}
    \gamma_{2} = \exp(\beta \cdot x_{1}^{2})
    \label{eqn:gama_2}
\end{equation}
According to the theory, $\beta$ should remain constant throughout all mixture composition. Rubingh's theory has been extensively applied to various surfactant systems due its simplicity, however model inconsistencies regarding the interaction parameter $\beta$ are reported~\citep{KumarShah2022,Grady2023,McLachlan2006,Sonu2013}. The estimation of the $CMC_\text{mix}$ can also be accomplished using numerous molecular-thermodynamic frameworks that have been proposed in the literature~~\citep{Shiloach1997,Shiloach1998,Iyer2014a,Srinivasan2005}. However, these models rely on multiple analytical approximations, are computationally intensive, have been applied on small number of systems and are not applicable on surfactants with a complex structure~\citep{Iyer2013}.

\par Machine learning (ML) models, specifically graph neural networks (GNNs), have shown very promising results in predicting the CMC of pure species for a wide variety of surfactants~\citep{Brozos2024,Qin2021,Moriarty2023,brozos2024graph}. GNNs treat every molecule as a graph, with atoms corresponding to nodes and bonds to edges. They learn to extract necessary structural information in an end-to-end framework and map it to the molecular property of interest. Recently, GNNs were expanded to binary mixtures for mixture properties such as activity coefficients and viscosity with very promising results~\citep{Rittig2022,Rittig2023,SanchezMedina2022,SanchezMedina2023,Qin2023,Bilodeau2023,Vermeire2021}. Yet, surfactant mixtures have not been investigated. Here, we extend GNNs to CMC predictions of surfactant mixtures at different temperatures.

\par We propose GNNs for CMC prediction of surfactant mixtures between surfactants of all surfactant classes. For this, we collect CMC data for 108 binary surfactant mixtures from the literature at multiple temperatures. The assembled data set consists of 599 data points. We enrich the assembled data set by concatenating pure species data from our previous work~\citep{Brozos2024}. Hence, the final data set contains 1,924 CMC values for both pure species and binary surfactant mixtures at various temperatures. To treat surfactant mixtures, we consider two GNN architectures: (i) a composition-based weighted linear summation of the molecular fingerprints of each pure species and (ii) a mixture graph accounting for inter-molecular interactions and hydrogen bonding. Since the performance of ML models can vary significantly depending on the test split considered~\citep{Yang2019,Zahrt2020}, we implement different splits/test scenarios for binary mixtures that are relevant for practical application of the model, e.g., predicting the $CMC_\text{mix}$ at different compositions or with new surfactant structures. In summary, we evaluate the accuracy of both GNN models for:
\begin{itemize}
    \item 4 different test scenarios that are based on potential practical applications of the model for binary surfactant mixtures
    \item 6 ternary surfactant mixtures collected from the literature
    \item 4 commercial surfactants, of which 2 are composed of two species (binary mixtures), 1 consists of three species (ternary mixture) and 1 consists of four species (quaternary mixture) for which the CMC is experimentally determined as part of work
    \item  mixtures of one commercial surfactant and sodium dodecyl sulfate (SDS) for which we experimentally determine the CMC as part of work.
\end{itemize}
\noindent In all these four subcases, the GNN models are not retrained; note in particular that the training was done solely on binary and pure species data and the GNNs need to extrapolate to ternary and quartenary mixtures.

\par We structure the article as follows: in Section~\ref{sec:dataset_all}, we provide a detailed description of the assembled data sets and the train-test splits. Section~\ref{sec:methods} refers to the architecture of the two GNN models and to the experimental procedures, while Section~\ref{sec:res_disc} offers an analysis and discussion of the model results. We summarize our findings in Section~\ref{sec:conclusion}. 

\par The final model predictions for the 6 ternary mixtures and for the 2 out of 4 test sets for binary mixtures are provided along with the source code, as open source in our GitHub repository. The model hyperparameters and architecture are presented in Section~\ref{sec:methods}. The training data set and the trained models remain the property of BASF and could be made available upon request.

\section{Data set}\label{sec:dataset_all}

\noindent We provide a general description of the data curated for binary 
and ternary mixtures (Sections~\ref{sec:dataset_overview} and~\ref{sec:ternary_description}), along with an analysis of the splitting procedure used (Section~\ref{sec:data_splits}). The newly generated experimental data for binary, ternary, and quaternary surfactant mixtures are presented in a later part of this work (cf. Section~\ref{sec:measurements}).

\subsection{Data set overview}\label{sec:dataset_overview}
\begin{figure}[htb]
    \centering
    \includegraphics[height =8.5cm, width = 11.5cm]{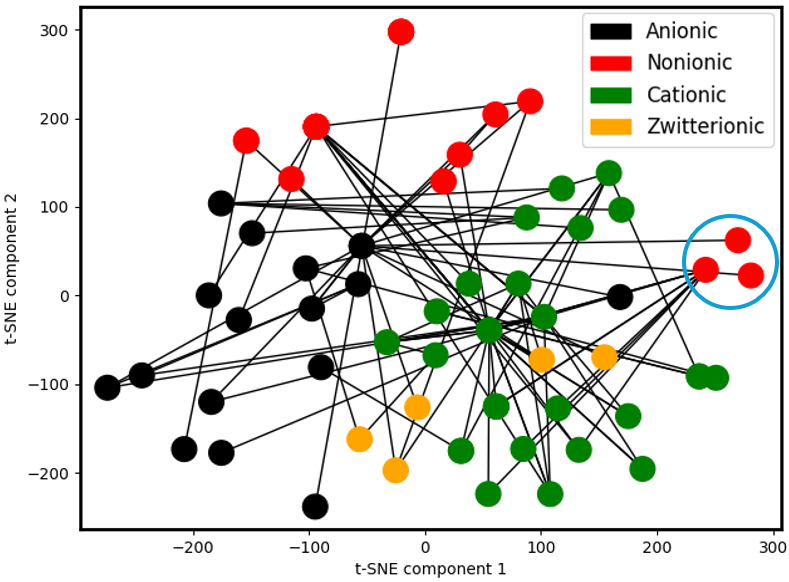}
    \caption{Mixture network of the curated data set. Each node represents a surfactant structure, and each edge the existence of a binary mixture between two surfactants. Correspondingly, the surfactants are also categorized based on their class. Each node is plotted on a 2D map obtained by applying t-SNE~\citep{Maaten2008VisualizingDU} on generated molecular fingerprints (ECFP\textunderscore10)~\citep{Rogers2010}. The surfactants are well distinguished in clusters based on their classes. An exception are the three n-alkyl-n-methylglucamides surfactants (Mega-8,-9 and -10) enclosed in the blue circle.}
    \label{fig:visualization_network}
\end{figure}

\noindent
We collect 108 binary surfactant mixtures at various temperatures from available literature sources~\citep{Zhang2004,Hines1997,Martin2010,Hierrezuelo2006,Prasad2006,Moulik1996,Treiner1992,udDin2009,Maiti2010}. The data set contains 515 experimental mixture points, i.e., data points where at least two surfactants coexist in the aqueous solution ($x_{1} \neq 0$ and $x_{2} \neq 0$), from a total of 68 pure species structures. A statistical overview of the 515 mixture points is provided in the Supporting Information (SI), Figure S1. We visualize the 108 binary surfactant mixtures as a mixture network, where each node represents one of the 68 surfactants, and each edge the existence of a binary mixture between two surfactants in Figure~\ref{fig:visualization_network}. For visualization of the structural similarity of the surfactants, we apply the t-distributed neighbor embedding (t-SNE)~\citep{Maaten2008VisualizingDU} on generated extended-connectivity fingerprints (ECFP\textunderscore10) for each pure species~\citep{Rogers2010}. 
As shown in Figure~\ref{fig:visualization_network}, the 68 surfactants are distinguished in clusters based on their classes. Some exceptions are also observed, with the most noticeable being the three nonionic surfactants inside the blue circle. All three of them are n-alkyl-n-methylglucamides, with different alkyl chain lengths (Mega-8,-9 and -10)~\citep{JeongSoo2004}. All three of them contain a nitrogen atom, as well as a branched alcohol compared to the other nonionic surfactants. As is evident in Figure~\ref{fig:visualization_network}, from the number of graph edges, some surfactants are well studied and are present in multiple mixtures, while others are present only in one mixture.

Based on the preceding discussion, one would expect the assembled data set to have a total size of 583, which will include the 515 mixture points and the 68 pure species data points. However, it is worth noting that some of the 68 pure species are present in mixtures at more than one temperature. For instance, SDS (a widely researched and used surfactant) is a component in mixtures at 4 different temperatures. Consequently, the herein assembled data set contains 4 distinct data points for pure SDS, i.e., the CMC at 4 different temperatures. The same holds true for other surfactants, resulting in a final data set size of 599 data points. Furthermore, to enrich molecular diversity and temperature-dependency information, we concatenate 1,377 CMC data points for pure species at various temperatures from our previous work to the binary mixture data set~\citep{Brozos2024}. In the combined data set, duplicate entries between the CMC values of the 68 newly collected surfactants as pure species and the 1,377 old data points arise, which are averaged. An example is given in the SI, Table S1. Overall, the final assembled data set consists of 1,924 data points. The minimum experimental temperature of our data is 0~$^\circ$C and the maximum is 90~$^\circ$C. 

\subsection{Data splits}\label{sec:data_splits}
\noindent
We split our data set in different ways into training and test set to evaluate the model for different practical settings. As the choice of the test set can impact how the model performance is interpreted, leading to over-confident results~\citep{Zahrt2020}, using different test scenarios is also beneficial to assess the robustness of the model. So we evaluate the performance of the model under different test scenarios. We implement 4 types of data set splitting: (i) composition interpolation (comp-inter), (ii) mixture compositions extrapolation (mix-comp-extra), (iii) mixture surfactant extrapolation (mix-surf-extra), (iv) mixture extrapolation (mix-extra) as described in the following paragraph. An overview of the 4 splits can be found in the SI, Table S2.

\par The \emph{comp-inter} test set contains mixture points of previously seen binary mixtures but at different compositions. To select the mixture points, we identify all binary surfactant mixtures with at least two different mixture compositions. Out of the 108 binary mixtures present in our data set, 96 of them fulfill this criteria. For each of them, a mixture point is randomly selected, removed from the training set, and assigned to the test set. By utilizing the comp-inter test set, we can assess model's performance in predicting new mixture points of a binary mixture for which measured other mixture points are readily available.

\par The \emph{mix-comp-extra} test set refers to binary surfactant mixtures, where the surfactants were seen during training in other surfactant mixture combinations. In other words, the training set includes the two surfactant structures of a binary mixture, either solely as pure species or as components of other mixtures. However, their combination (mixture) remains completely unseen. Two subsets, each containing 10 binary mixtures are randomly selected, consisting of 46 and 58 mixture points, respectively. Similarly to the comp-inter test set, the mix-comp-extra test set only contains mixture points. We can thereby evaluate whether the model is able to predict binary surfactant combinations for which experimental mixture data is not available yet.

\par The \emph{mix-surf-extra} test set expands the extrapolation character of the mix-comp-extra split by completely excluding one of the two surfactants of a binary mixture from the training data set. That is, one surfactant is not included in the training set, either as pure species or as a component of other mixtures, whereas the other surfactant is included, either as pure species or in other mixtures. This test scenario reflects the isolation/synthesis of a new surfactant structure, for which no previous measurements are available. To construct the mix-surf-extra test set, we select 4 representative surfactants, namely: n-decanoyl-n-methylglucamide (Mega-10), n-dodecyl-$\beta$-D-maltoside ($\beta$-C$_{12}$G$_{2}$), cetylpyridinium chloride (CPC) and SDS. Removing them from the training set simultaneously would result in a huge information loss, accounting to about 30 percent of the mixture points, and thus diminish model performance. Therefore, we decided to remove each of them (and corresponding mixtures) separately. Accordingly, we constructed 4 sub-test sets. The model performance on the mix-surf-extra test set is reported on the combination of the 4 sub-test sets (cf. Section~\ref{sec:res_disc}). For example, for Mega-10 there are 9 binary mixtures (45 mixture points) in the assembled data set at 30~$^\circ$C~\citep{Martin2010,Hierrezuelo2006}. The sub-test set will consist both the 45 mixture points and the Mega-10 as pure species at 30~$^\circ$C. The data points for pure Mega-10 at temperatures different than 30~$^\circ$C~\citep{Prasad2006} are excluded from both the training and the test set, since we already demonstrated GNN's ability to predict the temperature-dependent CMC of pure surfactants in our previous work~\citep{Brozos2024} and we herein focus on surfactant mixtures. Similarly, the other sub-test sets contains 3 binary mixtures (14 mixture points) at 25~$^\circ$C for $\beta$-C$_{12}$G$_{2}$~\citep{Zhang2004,Hines1997}, 6 binary mixtures (19 mixture points) at 25 and 30~$^\circ$C for CPC~\citep{Moulik1996,Treiner1992,udDin2009,Maiti2010}, and 15 binary mixtures (82 mixture points) at 22, 25, 30, 35 and 40~$^\circ$C for SDS~\citep{Haque1996,Hines1997,Hines1998,Maiti2010,LopezFontan2000}. We can thereby evaluate whether the model is able to predict binary mixture combinations of surfactants for which one of them has not been seen before during model training.

\par The mix-extra test set considers a scenario where none of the surfactants in a binary mixture are encountered during training, either as pure species or as components of other binary mixtures. In this scenario, the model has to predict CMCs of new/unseen surfactant structures, as well as binary mixtures of them. First, we identify 5 binary mixtures composed by 6 surfactants (2 anionic and 4 nonionic) that fulfill the criteria described above~\citep{Huang2017,Ren2014,Haque1996}. To enhance structural complexity and variety of the mix-extra test set, we further assign 2 binary mixtures composed by 3 cationic surfactants~\citep{Treiner1992}. However, one of the 3 cationic surfactants (cetalkonium chloride) is present in a mixture with CPC~\citep{Treiner1992}, which as described above, is widely present in the data set. We discard this mixture only for the mix-extra split. Thus, the training set contains 100 binary mixtures (and the data points of pure species) and the test set contains 7 binary mixtures. Hence, we can evaluate whether the model is able to predict binary mixtures of which both surfactants have never seen by the model before.
 
\subsection{Ternary mixtures from literature sources}\label{sec:ternary_description}
We further assemble a small external data set containing 6 ternary mixtures from literature sources~\citep{Moulik2021,Huang2019}. The ternary data set contains 16 mixtures points at 25 and 30~$^\circ$C, composed from 8 pure species structures. 
All 8 surfactant structures are present in the collected data set described in Section~\ref{sec:dataset_overview}. Here, we investigate whether the model can effectively generalize to ternary mixtures containing species for which CMC data are readily available as pure species and in binary mixtures. Note that the ternary mixtures are not used for  training. Additionally, we underscore the lack of CMC data for ternary surfactant mixtures in literature, likely due to the combinatorial increase of required measurements and resources incurred to do such measurements.

\section{Methods}\label{sec:methods}
\noindent
We begin this section by describing the composition-based weighted linear summation GNN architecture (Section~\ref{sec:gnn}). We then introduce the mixture graph framework; the second GNN architecture (Section~\ref{sec:mixture_graph}) and describe the hyperparameters and model implementation (Section~\ref{sec:hyperparameters}). Lastly, we provide a description of the experimental material and procedures (Section~\ref{sec:measurements}).

\subsection{Weighted sum GNNs for surfactant mixtures}\label{sec:gnn}
\begin{figure}
    
    \includegraphics[height = 4. cm, width = 16.cm]{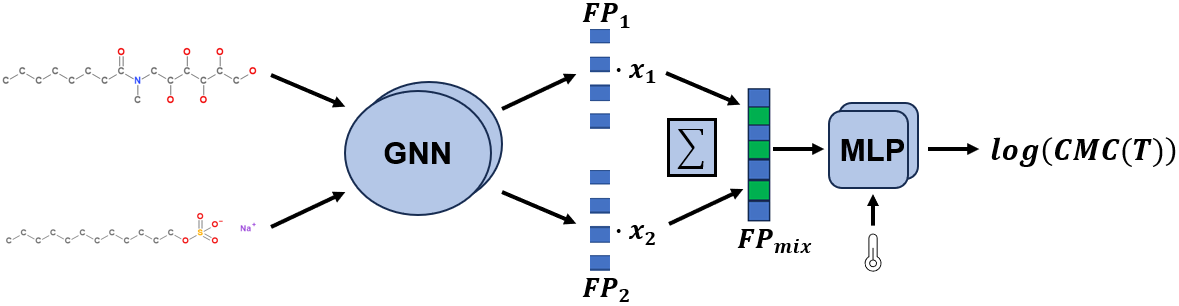}
    \caption{Schematic representation of the WS-GNN architecture for a binary mixture.}
    \label{fig:ws_architecture}
\end{figure}

\noindent
We develop a composition-based weighted linear GNN model for predicting the temperature-dependent CMC of surfactant mixtures.
The concept of GNN models has been described in detail in several literature sources, see Refs.~\citep{Yang2019,Zhou2020,9046288,Reiser2022,Khemani2024,hamilton2018inductive,Schweidtmann2020,Rittig2022a}.
We build on our GNN model for predicting the temperature-dependent CMC of pure species from our previous work~\citep{Brozos2024}. 
Here, we adapt the architecture for mixtures. 
Each surfactant molecule in the mixture is represented as a graph $G_{i}= (V,E)$, where $V$ are the vertices (atoms), $E$ are the edges (bonds) between nodes, $i \in \mathcal{N}$ is a component of the surfactant mixture and $ \mathcal{N}$ is the total number of mixture components. The GNN first encodes structural information from pure species through graph convolutions and a pooling step into a molecular fingerprint (FP), which is denoted as $FP_{i}$. 
To treat mixtures, we propose the weighted linear summation of the learned FPs, i.e., the mole fraction $x_{i}$ is used as weight. 
Hence, the mixture fingerprint $FP_\text{mix}$ is calculated through Eq.~\ref{eqn:mixture_fingerprint}. 
This model architecture enables also considering pure species, i.e., $x_{2}$ = 0 in the case of binary mixtures.
It further ensures permutation invariance with respect to the representation order of the surfactants within a mixture: if the order of the surfactants is changed, the $FP_\text{mix}$ remains the same.
We denote this architecture as \textbf{WS-GNN} (weighted sum GNN) and a schematic representation for an exemplary binary mixture is given in Figure~\ref{fig:ws_architecture}. Note that Eq.~\ref{eqn:mixture_fingerprint} does not explicitly encode the inverse relationship between $CMC_\text{mix}$ and $CMC_\text{i}$ described in Eqs.~\ref{eqn:clint_method},~\ref{eqn:activity_coef}. The reason for this is to allow greater flexibility during training, and since Eqs.~\ref{eqn:clint_method},~\ref{eqn:activity_coef} are only semi-empirical, they may not always hold true, as was found by many authors~\citep{Huang2017,Zhang2004,MisselynBauduin2000}.

\begin{equation}
     FP_\text{mix} = \sum_{i \in N} x_{i} \cdot FP{_i}
    \label{eqn:mixture_fingerprint}
\end{equation}

The $FP_\text{mix}$ is mapped to the temperature-dependent CMC through a standard multi-layer perceptron (MLP). The non-linearity introduced in the MLP allows for the capture of the non-linear surfactant mixing behavior. The temperature is concatenated to the first hidden layer, similar to our previous work~\citep{Brozos2024}, where more detailed information regarding the MLP architecture and temperature dependency can be found.

\subsection{GNNs with mixture graph}\label{sec:mixture_graph}
Alternatively, to capture molecular interactions in mixtures, more advanced geometries that consider hydrogen bonding information have been recently proposed~\citep{SanchezMedina2023,Qin2023}. Here, the construction of a new graph is proposed, where nodes represent the pure species of each mixture and edges inter- and intramolecular interactions~\citep{SanchezMedina2023,Qin2023}. The pure species fingerprints are used as node feature vectors and number of hydrogen acceptors and donors as edge features~\citep{SanchezMedina2023,Qin2023}. Afterwards, the mixture graph is passed into a GNN layer to account for intermolecular interactions~\citep{SanchezMedina2023,Qin2023}.

\begin{figure}[b!]
	\includegraphics[height = 3.8 cm, width = 16.cm]{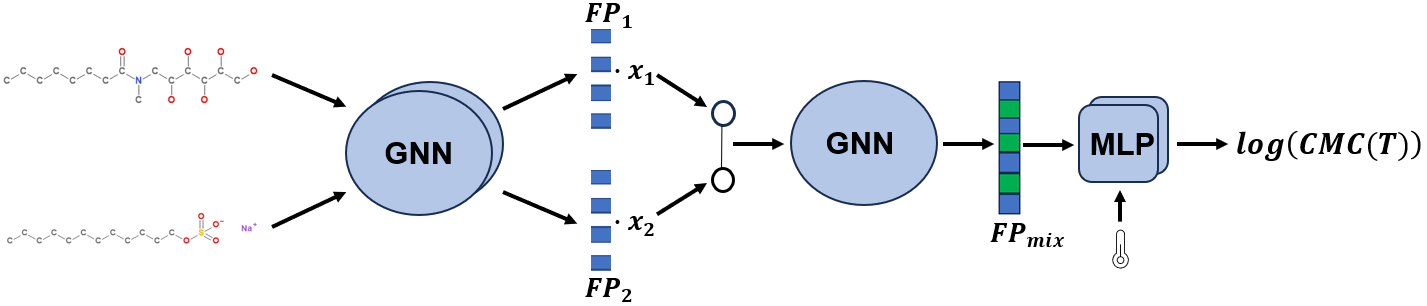}
	\caption{Schematic representation of the MG-GNN architecture for binary mixtures.}
	\label{fig:mg_architecture}
\end{figure}

\par We adapt the proposed mixture graphs architectures by firstly multiplying the composition $x_{i}$ of surfactant $i$ with the corresponding $FP{_i}$. The composition-adjusted  $FP{_i}$ are utilized as node feature vectors of the mixture graph, as proposed in previous works~\citep{SanchezMedina2023,Qin2023}. For edge features, we also consider the number of hydrogen bond acceptors and donors, but we calculate them through Lipinski's rule of five~\citep{Lipinski2001}, as we observed a slightly higher model accuracy. Then, the mixture graph is passed into a graph convolutional layer to account for surfactant interactions~\citep{SanchezMedina2023,Qin2023}. We use the GINE-operator~\citep{xu2019powerful,hu2020strategies} as a graph convolutional layer. To extract the $FP_\text{mix}$, a summation pooling layer is added, as was also proposed in our recent work~\citep{rittig2024thermodynamics}. Analogously to the WS-GNN, the pooling layer ensures permutation invariance for the surfactant order within the mixture. We denote this architecture as \textbf{MG-GNN} (mixture graph GNN) and illustrate it for an exemplary binary mixture in Figure~\ref{fig:mg_architecture}.

\subsection{Implementation, Hyperparameter Tuning and Ensemble learning}\label{sec:hyperparameters}
\noindent
All models are implemented in PyTorch Geometric (PyG)~\citep{Fey2019}. We perform hyperparameter tuning based on the WS-GNN architecture (cf. Section~\ref{sec:gnn}). 
Specifically, we train the GNN model on 20 different, randomly selected validation sets, similar to our previous works~\citep{Brozos2024, brozos2024graph,Rittig2022}. 
The size of the validation set is kept constant at 385 points, which represents 20\% of the whole data set size. We use the root mean square error (RMSE) on the the comp-inter split for hyperparameter tuning, which are provided in the SI, Table S3. 

\par To improve the predictive capabilities of ML models, ensemble learning is a widely employed. The models trained on different seeded validation sets may produce noisy results. By averaging out the predictions of all them, robust and generalized predictions are obtained~\citep{Breiman1996, Dietterich2000, Ganaie2022}. We average the predictions of the 20 different trained models for each split type mentioned in Section~\ref{sec:data_splits}, and report only the prediction accuracy of the ensemble of GNNs~\citep{Brozos2024,Rittig2022}. We further apply ensemble learning when combining the two developed GNN models, described in Sections~\ref{sec:gnn} and~\ref{sec:mixture_graph}. That is, the predictions from the two proposed GNNs architectures (WS-GNN and MG-GNN) are averaged.

\begin{figure}[b!]
	\centering
	\includegraphics[height = 5 cm, width = 6 cm]{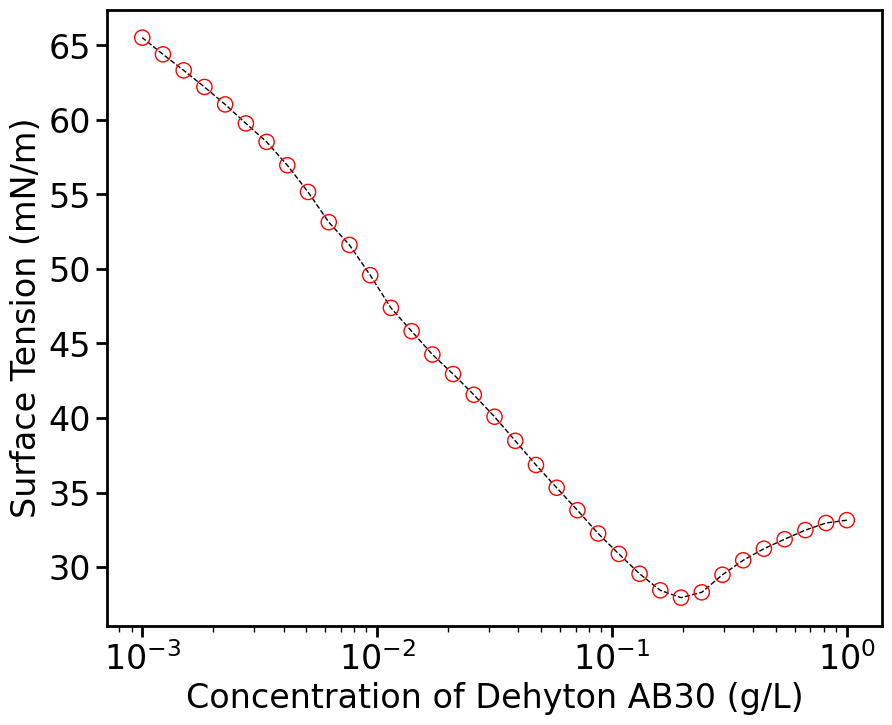}
	\caption{Surface tension measurement of D-AB30 at 23~$^\circ$C.}
	\label{fig:ab30_sft}
\end{figure}

\subsection{Materials and Surface tension measurements}\label{sec:measurements}
As industrial grade surfactants are usually composed from more than one species, we aim to identify such examples and validate the accuracy of our model. The commercial surfactant Dehyton\textsuperscript{\textregistered} AB 30 (D-AB30) is provided by BASF and was recently used in a commercially formulation study~\citep{Cao2021}. A gas chromatography (GS) analysis is also provided from the supplier. The main two species ($\geq$ 98\%) are lauryl and myristyl betaine. Therefore, the product can be well categorized as a binary surfactant mixture between two zwitterionic surfactants. We note that both pure species exist in our collected data set (cf. Section~\ref{sec:dataset_overview}) but no mixture between two zwitterionic surfactants is available. Furthermore, highly purified unary SDS ($\geq$ 99\%) was bought from Sigma Aldrich and used as received. Mixtures between SDS and D-AB30 at different mole fractions were prepared and measured. 
Since D-AB30 is a binary mixture, the prepared mixtures are considered ternary ones. Three more commercial surfactants, namely Sulfopon\textsuperscript{\textregistered} 1214 G ($\geq$ 96\%), Texapon\textsuperscript{\textregistered} V 95 G ($\geq$ 96\%) and Texapon\textsuperscript{\textregistered} K 30 UP ($\geq$ 98\%) were provided by BASF, together with a GC analysis. The Sulfopon\textsuperscript{\textregistered} 1214 G (S-1214G) is composed from two species, namely SDS and sodium tetradecyl sulfate (STS)  while the Texapon\textsuperscript{\textregistered} V 95 G (T-V95G) consists three species, namely SDS, STS and sodium hexadecyl sulfate. We note, that no mixture data between these species exist in our collected data sets (cf. Sections~\ref{sec:dataset_overview} and~\ref{sec:ternary_description}). The Texapon\textsuperscript{\textregistered} K 30 UP (T-K30UP) consists four species, namely SDS, STS, sodium hexadecyl sulfate and sodium octadecyl sulfate.

\par Surface tension measurements were performed at 23~$^\circ$C using the Force Tensiometer – K100 (Krüss, Germany) and the Wilhelmy plate method with a standardized platinum-iridium plate. The surfactant solutions were freshly prepared in deionized water and adjusted to pH 4.9 prior to the measurements. The surface tension was plotted against the logarithm of the surfactant concentration to determine the CMC. To ensure reproducibility, the measurements were repeated three times. The surface tension measurement of D-AB30 is illustrated in Figure~\ref{fig:ab30_sft}.

\section{Results and discussion}\label{sec:res_disc}
\noindent 
We begin this section by discussing the results of the two GNN models on the 4 different test scenarios for binary mixtures (Section~\ref{sec:pred_perf}). Afterwards, we analyze the model performance through different surfactant class combinations (Section~\ref{sec:surf_mix_clas}) and the GNN model applicability on ternary mixtures (Section~\ref{sec:pred_ternaries}). We then compare model predictions with experimental values for commercial surfactants and mixtures of them (Section~\ref{sec:exp_val}). We conclude this section by comparing the GNN models to a semi-empirical model (Section~\ref{sec:semi_empirical}).

\subsection{Predictive performance on binary surfactant mixtures}\label{sec:pred_perf}
\begin{figure}
	\centering
	\captionsetup[subfloat]{font=small}
	\subfloat[Comp-inter test set]{\includegraphics[height =7.5 cm, width = 8.cm]{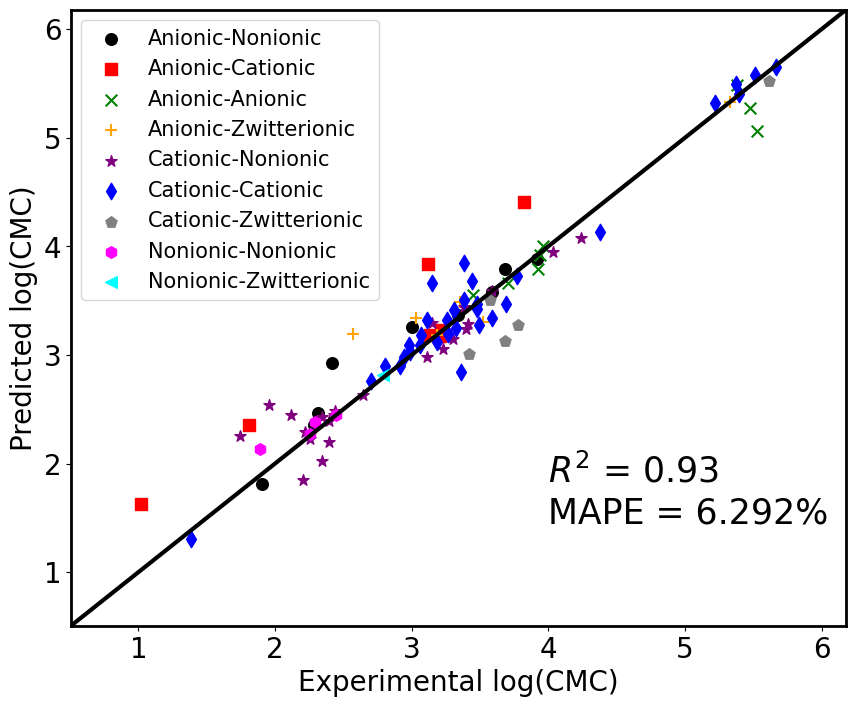}}   
	\subfloat[Mix-comp-extra test set]{\includegraphics[height =7.5 cm, width = 8.0cm]{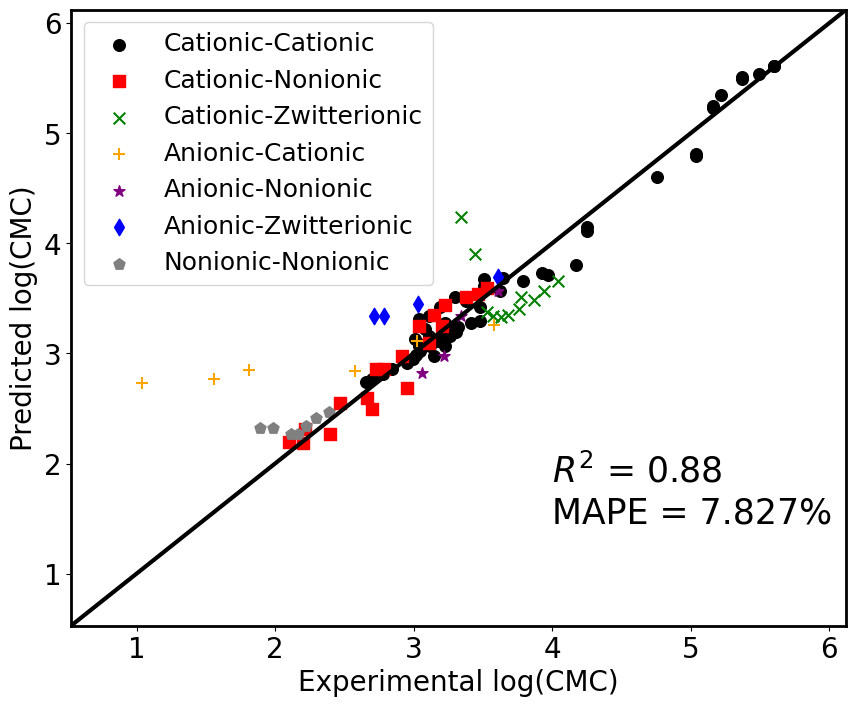}}  
	\hfill
	\subfloat[Mix-surf-extra test set]{\includegraphics[height =7.5 cm, width = 8.cm]{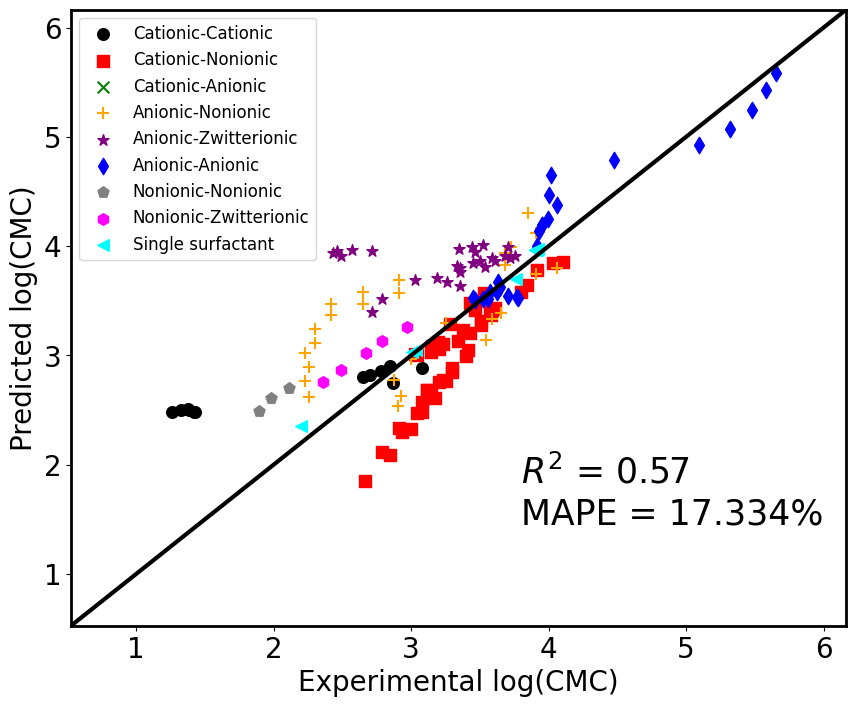}}   
	\subfloat[Mix-extra test set]{\includegraphics[height =7.5 cm, width = 8.cm]{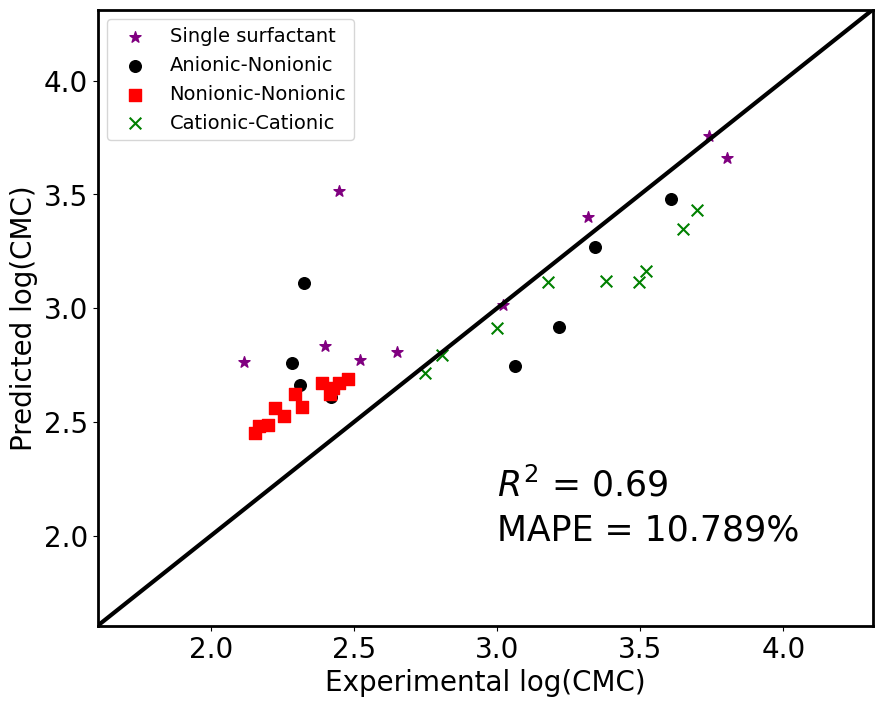}}  
	\caption{Parity plots on the 4 test sets. The predictions are made by the combined GNN model. The data points are highlighted with different colors and markers based on the classes of the two mixture species. The logarithm is applied to CMC in $\mu$M (base 10). }
	\label{fig:parity_plot}
\end{figure}

\noindent
In Table~\ref{tab:error_matrices}, an overview of the prediction accuracy of the two developed GNN architectures is presented for the 4 different test sets (cf. Section~\ref{sec:data_splits}). We report 4 different error metrics for each test set.
We scale the CMC (µM) values using a (based on 10) logarithmic scale and hence all error metrics refer to the $log(\text{CMC})$. From Table~\ref{tab:error_matrices}, we observe that the best performing architecture varies for each split, i.e., no clear best-performing GNN model can be identified. Therefore, we consider averaging their predictions (cf. Section~\ref{sec:hyperparameters}) and denote the model as \textbf{combined}. Another motivation is that the combined framework leverages learning from both architectures and offers a more flexible and robust framework. The combined results are reported in Table~\ref{tab:error_matrices} and parity plots on the 4 test scenarios are illustrated in Figure~\ref{fig:parity_plot}.

\begin{table}[b!]
	\caption{Summary of the prediction accuracy of the ensemble GNNs models, i.e., WS-GNN and MG-GNN architectures, as well as of the combination of the two GNN architectures on the 4 test scenarios. MAE = mean absolute error, MAPE = mean absolute percentage error (unit \%).}
	\centering
		\begin{tabular}{c   c | c  c  c  c  c}
			Model &  & comp-inter & mix-comp-extra & mix-surf-extra & mix-extra \\ 
			\hline
			\cline{3-6}
			WS-GNN & \multirow{3}{*}{RMSE} & 0.264 & \textbf{0.302} & \textbf{0.507} & 0.441 \\
			MG-GNN &  & 0.251 & 0.334 & 0.665 & \textbf{0.333} \\
			Combined & & \textbf{0.249} & 0.313 & 0.551 & 0.344 \\ 
			\hline
			WS-GNN  & \multirow{3}{*}{MAE} &  0.183 & \textbf{0.182} & \textbf{0.382} & 0.366\\
			MG-GNN &  & 0.178 & 0.215 & 0.486 & \textbf{0.215} \\
			Combined & & \textbf{0.171} & 0.196 & 0.406 & 0.274 \\ 
			\hline
			WS-GNN& \multirow{3}{*}{MAPE} &  6.831 & \textbf{7.352} & \textbf{15.916} & 14.758\\
			MG-GNN &  & 6.344 & 8.452 &20.336 &\textbf{8.2} \\
			Combined & & \textbf{6.292} & 7.827 & 17.334 & 10.789 \\ 
			\hline
			WS-GNN & \multirow{3}{*}{R$^2$} & 0.93 & \textbf{0.89} & \textbf{0.62}  & 0.54 \\ 
			MG-GNN &   & 0.93 & 0.86 & 0.43 & 0.64 \\
			Combined & & \textbf{0.93} & 0.88 & 0.57 & \textbf{0.69} \\ 
		\end{tabular}
	\label{tab:error_matrices}
\end{table}

In the comp-inter split, the combined model outperforms the best performing MG-GNN model, as indicated by the slight decrease in RMSE from 0.251 to 0.249 (cf. Table~\ref{tab:error_matrices}). In the mix-comp-extra split, it yields an RMSE of 0.313, which is very close to the lowest RMSE (0.302) achieved by the WS-GNN model. In both test scenarios, the RMSE is similar to that reported from previous accurate GNN models applied to pure species~\citep{Brozos2024,Qin2021,Moriarty2023,brozos2024graph}. In the parity plots of the two test sets (Figures~\ref{fig:parity_plot}a,b), the majority of the points lie very close to the diagonal, as can also be seen from the high R$^{2}$ values, 0.93 and 0.88 respectively. Therefore, the model can be used to provide highly accurate predictions for these two test scenarios. 

\par In the mix-surf-extra split, the combined model yields an RMSE of 0.551, which is also close to the RMSE (0.507) of the WS-GNN model. Exclusion of the pure species of a binary mixture from the training set, drastically reduces the model performance, as a significantly higher error is exhibited in the mix-surf-extra test set. However, most of the predictions in Figure~\ref{fig:parity_plot}c also lie close to the diagonal. Here, the noticeable outliers significantly increase the RMSE. Furthermore, the combined model exhibits an RMSE of 0.344 on the mix-extra test, very similar to the lowest RMSE (0.333) achieved by the MG-GNN model. The predictions lie closer to the diagonal as in the mix-surf-extra test set, which is illustrated from the higher R$^{2}$ value, 0.69 compared to 0.57. The results of extrapolation to new surfactant structures and binary mixtures indicate that model performance depends on the new, unseen structures introduced to the model. In essence, a high fraction of the predictions are very accurate (indicated from the parity plots), but outliers with high deviation are more common than in the comp-inter and mix-comp-extra splits. Therefore, the model should be used with increased awareness in such extrapolation scenarios.

\subsection{Predictive performance per surfactant mixtures classes}\label{sec:surf_mix_clas}
\noindent

In this subsection, we provide an analysis of model results per surfactant class combinations. We categorize the binary surfactant mixtures based on the classes of the two species and we identify descriptive examples in each test set split. The predictions are made with the combined GNN model. All the comparisons refer again to the logarithmic scale. For the interested reader, selective examples with absolute CMC values are illustrated in the SI, Figure S2.

 \begin{figure}[htbp]
    \centering
    \captionsetup[subfloat]{font=small}
    \subfloat[Mix-comp-extra test set]{\includegraphics[height =6. cm, width = 6.6cm]{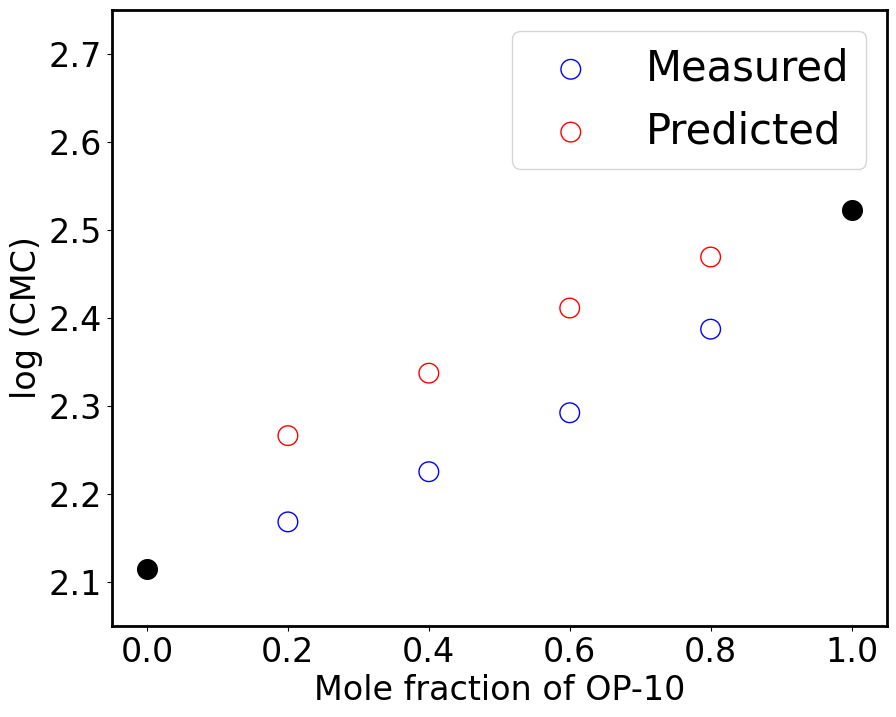}}    
    \subfloat[mix-extra test set]{\includegraphics[height =6. cm, width = 6.6cm]{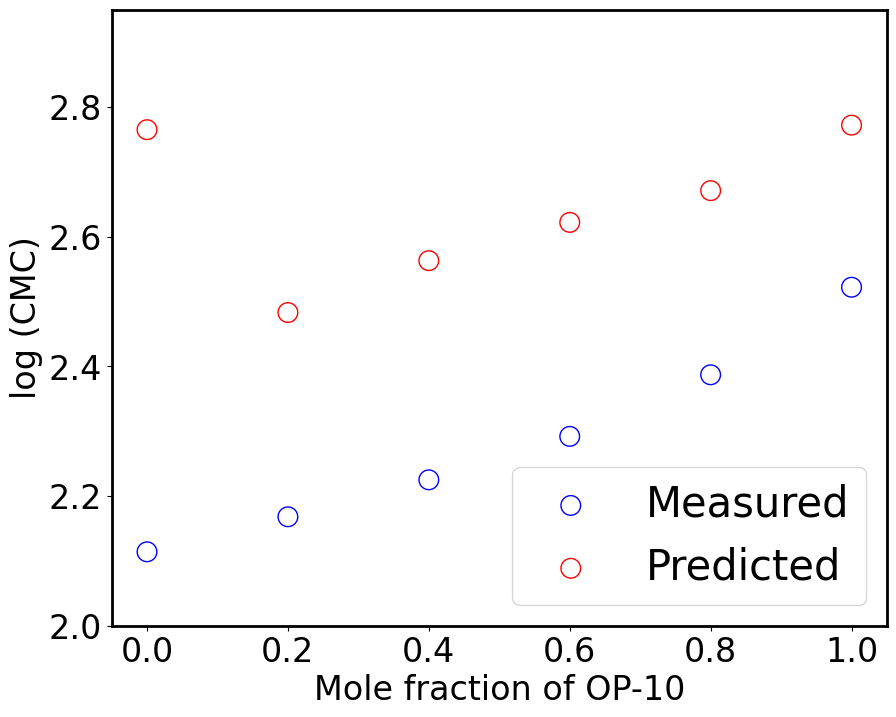}} 
    \vspace{5mm}
    \hfil
    \subfloat[Mix-comp-extra test set]{\includegraphics[height =6. cm, width = 6.6cm]{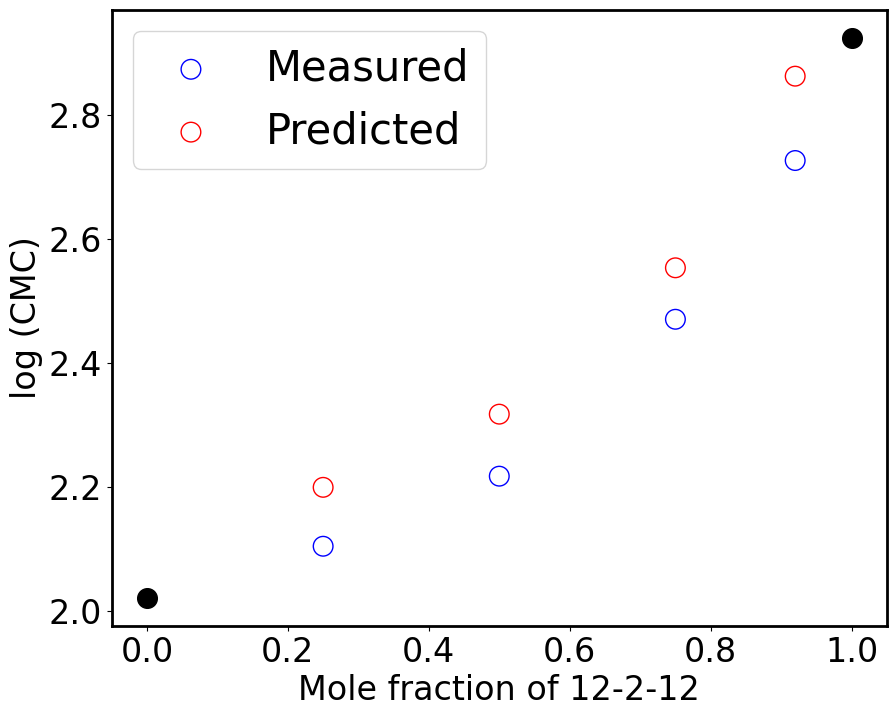}}   
    \subfloat[mix-surf-extra test set]{\includegraphics[height =6. cm, width = 6.6cm]{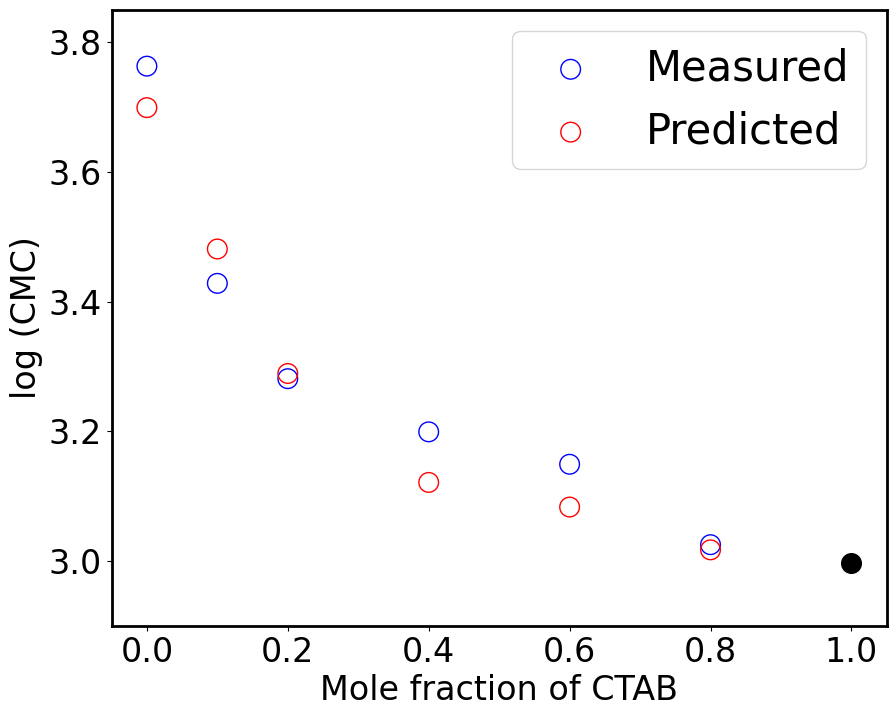}} 
    \caption{GNN predictions versus experimental data on different surfactant mixtures classes are present in three test scenarios. Panel (a) and (b) refer to a mixture between nonionic OP-10 and OP-4~\citep{Huang2017}. Panels (c) and (d) refer to mixtures between cationic and nonionic surfactants, namely 12-2-12/C$_{12}$E$_{8}$ and CTAB/Mega-10~\citep{Alargova2001, Hierrezuelo2006}. Black filled circles represent data points present in the training set.}
    \label{fig:nonionic_nonionic}
\end{figure}

\subsubsection{Mixtures with nonionic surfactants}
Nonionic surfactants typically show ideal micelle formation when mixed with other nonionic surfactants and synergistic behavior when mixed with ionic surfactants (cf. Section~\ref{sec:introduction}). A nonionic/nonionic mixture between octylphenol polyoxyethylene ether (OP-10) and OP-4 at 25~$^\circ$C is presented in Figures~\ref{fig:nonionic_nonionic}a,b~\citep{Huang2017} in two different test scenarios. 
In the case of mix-comp-extra test set, the combined GNN model accurately predicts the trend and the $CMC_\text{mix}$ in all mole fractions. Higher errors are observed for the mix-extra test set. The model significantly overestimates the CMC of pure OP-4, thus leading to overestimated $CMC_\text{mix}$ predictions in lower OP-10 mole fractions. Figure~\ref{fig:nonionic_nonionic}c refers to the mixture of dimethylene-1,2-bis(dodecyldimethylammonium bromide) (12-2-12), a dimeric cationic surfactant, and C$_{12}$E$_{8}$~\citep{Alargova2001} and Figure~\ref{fig:nonionic_nonionic}d represents the mixture of cetyltrimethylammonium bromide (CTAB), a cationic surfactant, and Mega-10~\citep{Hierrezuelo2006}. Hence, both cases belong to cationic/ nonionic combinations. In both cases, highly accurate predictions are demonstrated throughout all mole fractions. To summarize, in all 4 examples the trend is well captured except for mixture (b) at composition $x_{1}=0$.

\subsubsection{Ionic - Ionic mixtures}
\begin{figure}
    \captionsetup[subfloat]{font=small}
    \subfloat[Anionic / Cationic]{\includegraphics[height =5.3 cm, width = 5.3cm]{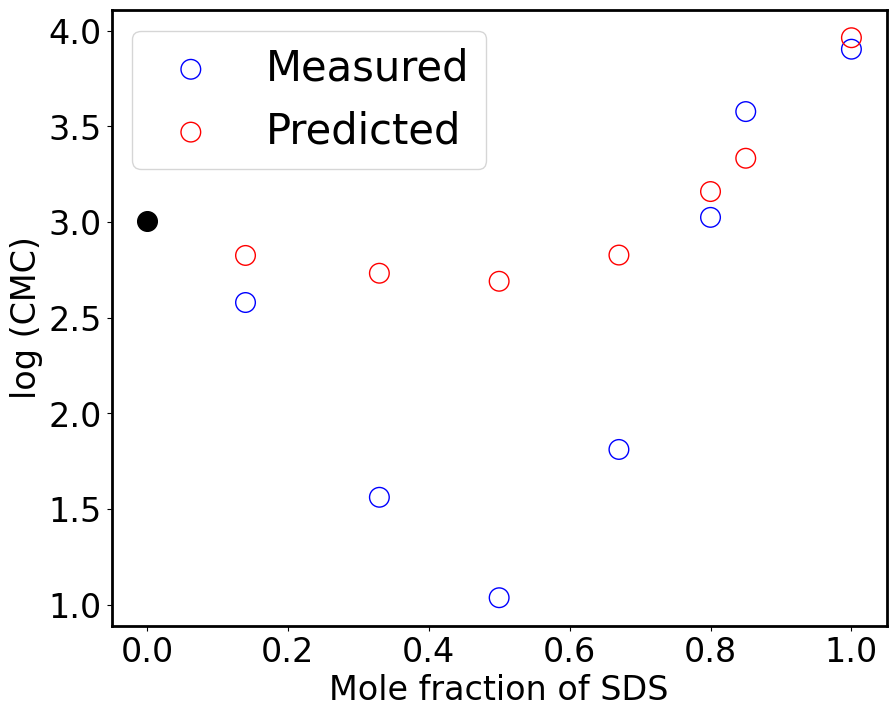}}
    \subfloat[Anionic / Anionic]{\includegraphics[height =5.3 cm, width = 5.3cm]{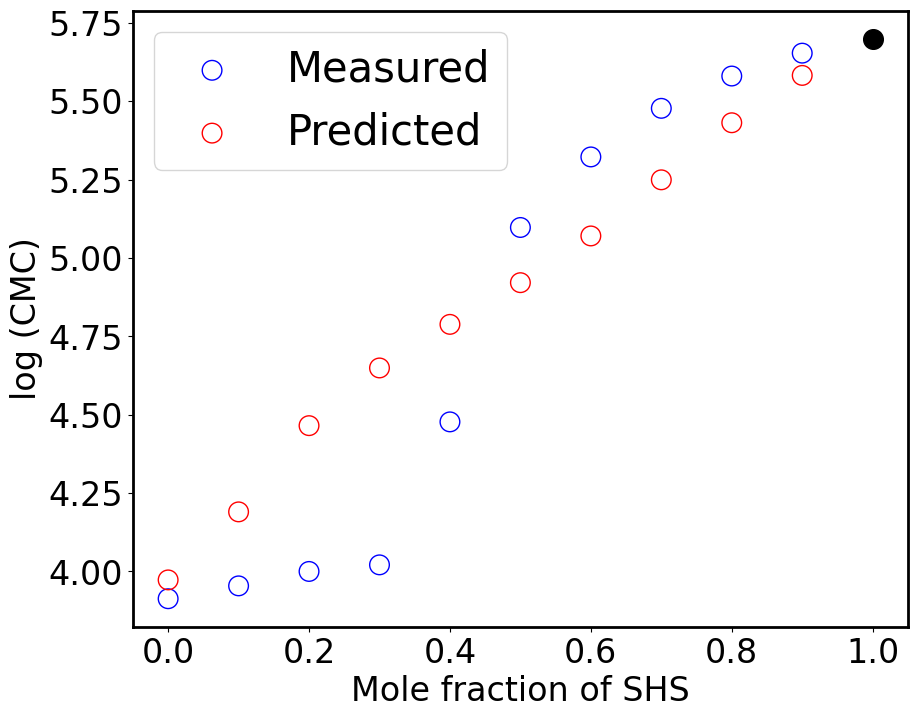}}
    \subfloat[Cationic / Cationic]{\includegraphics[height =5.3 cm, width = 5.3cm]{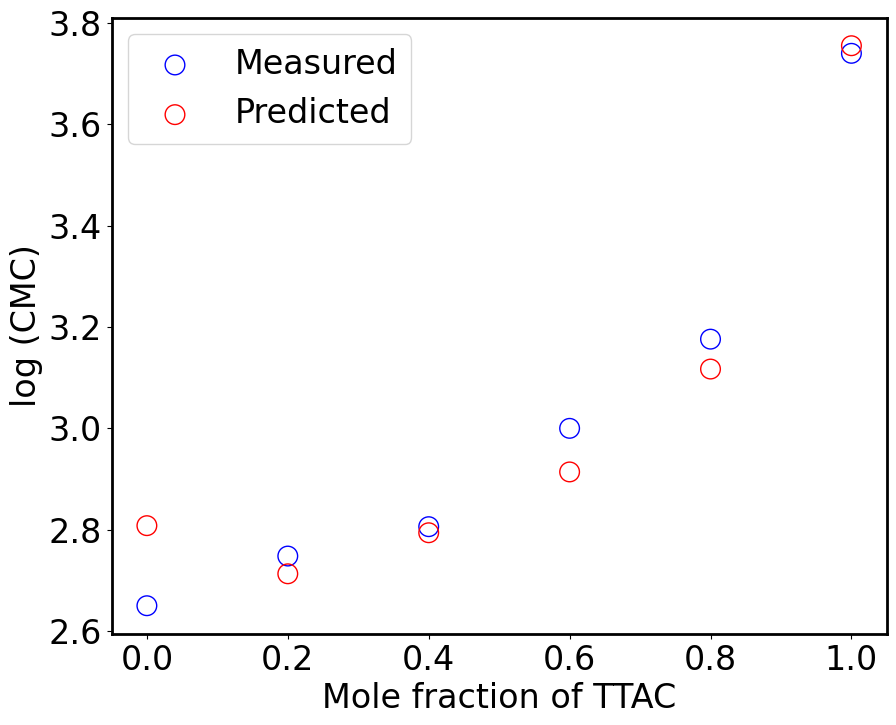}}   
    \caption{GNN predictions versus experimental data on three different surfactant mixture combinations present on the mix-surf-extra  and mix-extra test sets. Panel (a) refers to a mixture between SDS and CPC~\citep{Maiti2010}, panel (b) to a mixtures between SHS and SDS~\citep{LopezFontan2000} and panel (c) between TTAC and benzyldimethyltetradecylammonium chloride~\citep{Treiner1992}. Black filled circles represent data points present in the training set.}
    \label{fig:ionic_comb}
\end{figure}

Mixtures between ionic surfactants can exist in three different combinations, namely anionic/cationic (opposite charge of the head group), anionic/anionic and cationic/cationic (similar head group charge). Here, we investigate all three of them. In Figure~\ref{fig:ionic_comb} one exemplary mixture for each combination is presented. Figure~\ref{fig:ionic_comb}a refers to a mixture between SDS and CPC at 25~$^\circ$C~\citep{Maiti2010}, while Figure~\ref{fig:ionic_comb}b to a mixture between sodium hexyl sulfate (SHS) and SDS at 35~$^\circ$C~\citep{LopezFontan2000}. Both mixtures are extracted from the mix-surf-extra test set. Figure~\ref{fig:ionic_comb}c refers to a mixture between tetradecyltrimethylammonium chloride (TTAC) and benzyldimethyltetradecylammonium chloride at 25~$^\circ$C present in the mix-extra test set~\citep{Treiner1992}. The mixture containing two anionic species shows antagonistic behavior, which is accurately captured by the model. However, the model's sensitivity to mole fraction variances could be further improved. It is important to note that due to the removal of SDS from the training set, only three mixtures between anionic surfactants were present in the training data. Furthermore, in the anionic/cationic mixture, the strong synergism at mole fraction 0.5 is not fully captured, the combined GNN model accurately predicts the existence and point of synergism as well as the order of magnitude of $CMC_\text{mix}$ at other mixture compositions. The synergistic behavior between two cationic surfactants is accurately captured by the combined GNN model, as evident in Figure~\ref{fig:ionic_comb}c. The predicted $CMC_\text{mix}$ values at all mole fractions match the measured ones.
 
\begin{figure}[b!]
    \subfloat[Mix-comp-extra test set]{\includegraphics[height =5.3 cm, width = 5.3cm]{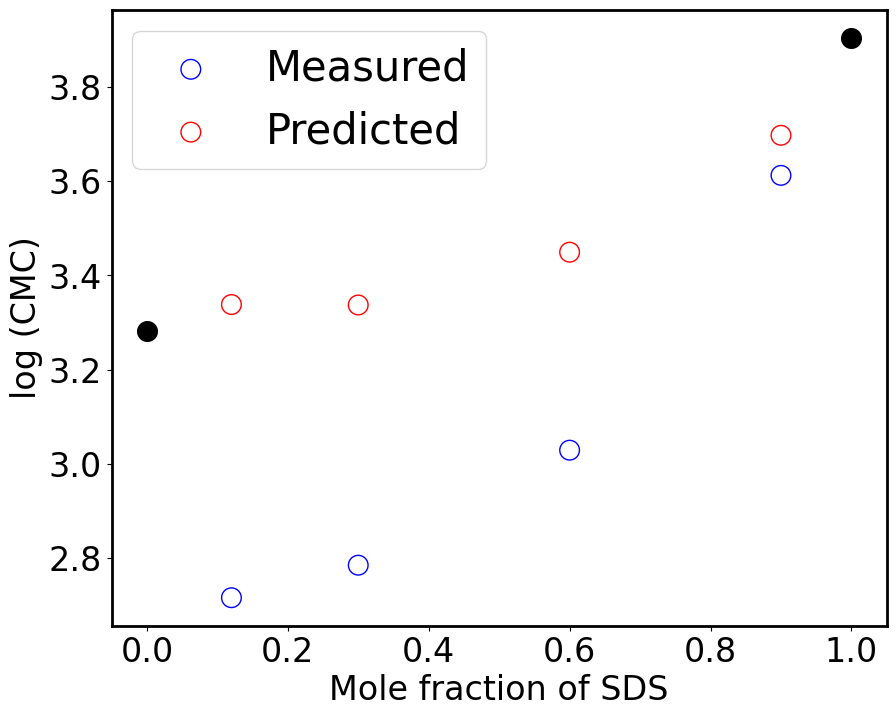}}   
    \subfloat[Mix-comp-extra test set]{\includegraphics[height =5.3 cm, width = 5.3cm]{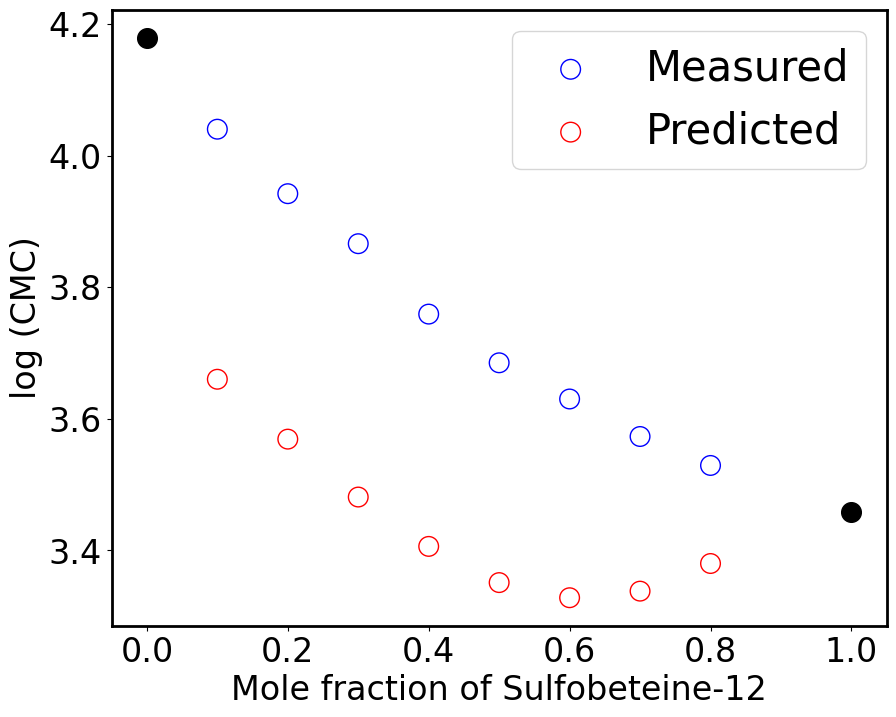}} 
    \subfloat[mix-surf-extra test set]
    {\includegraphics[height =5.3 cm, width = 5.3cm]{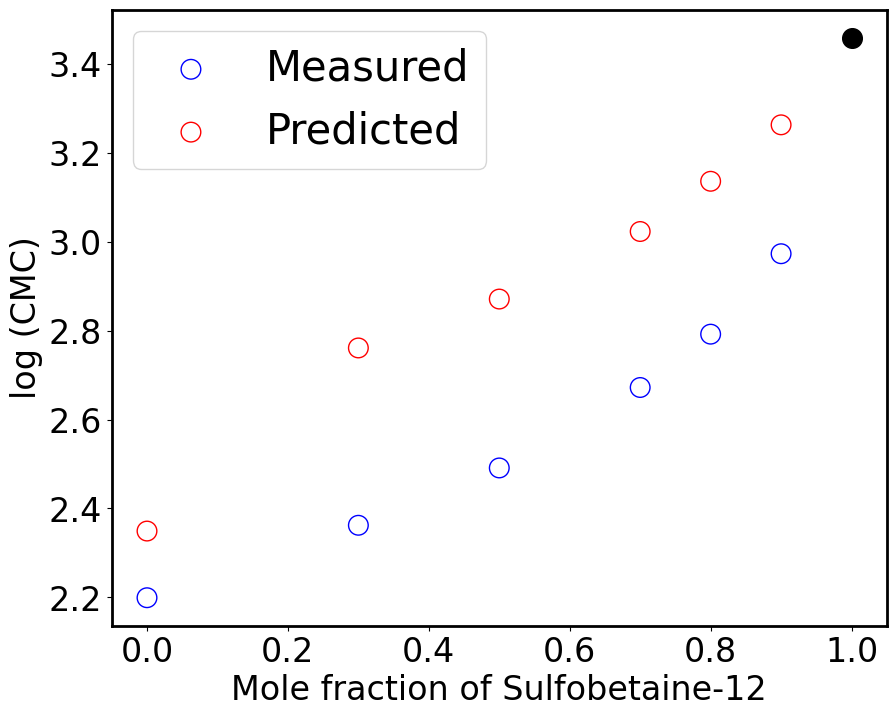}}   
    \caption{GNN predictions versus experimental data on mixtures with one zwitterionic species. Panel (a) refers to a mixture between SDS and LDAO~\citep{Bakshi1993}, panel (b) S-12 and DTAB~\citep{McLachlan2006} and panel (c) between S-12 and C$_{12}$G$_{2}$~\citep{Hines1997}. Black filled circles represent data points present in the training set.}
    \label{fig:amphoteric}
\end{figure}
\subsubsection{Mixtures with zwitterionics}
Binary mixtures, where at least one surfactant is zwitterionic, can exhibit nonideal mixed micelles and are highly pH sensitive. However, research on them has remained limited and hence only minor examples are present on our test sets. An anionic/zwitterionic mixture between SDS and lauramine oxide (LDAO) at 25~$^\circ$C is shown in Figure~\ref{fig:amphoteric}a~\citep{Bakshi1993}, while a cationic/zwitterionic mixture between dodecyltrimethylammonium bromide (DTAB) and sulfobeteine-12 (S-12) is presented in Figure~\ref{fig:amphoteric}b~\citep{McLachlan2006}. Both mixtures are present on the mix-comp-extra test set. A further example, namely a mixture between S-12 and C$_{12}$G$_{2}$ at 25~$^\circ$C is provided in Figure~\ref{fig:amphoteric}c~\citep{Hines1997}. For the first mixture, the model fails to capture the synergism of the system and significant deviations from the measured $CMC_\text{mix}$ are also observed. Similar behavior is observed at the mixture between DTAB and S-12, but the model exhibits accurate sensitivity regarding the mole fraction variations. For the mixture of S-12 and C$_{12}$G$_{2}$ good agreement between predicted CMCs and measured are observed. All over the combined model is able to capture the trend and provide good predictions.

\subsection{CMC predictions on ternary mixtures from literature sources}\label{sec:pred_ternaries}
\noindent

We adapt both GNN architectures to predict ternary mixtures described in Section~\ref{sec:ternary_description}, while training them exclusively on binary mixtures and pure species data. Here, we are interested in whether GNN models trained on CMC values of binary mixtures and pure species can accurately scale up to ternary mixtures with no further information. We use the trained WS-GNN and MG-GNN models from the comp-inter split to perform predictions on the 6 ternary mixtures described in Section~\ref{sec:ternary_description}. The results are given in Table~\ref{tab:ternary_results}. The WS-GNN model performs fairly good with and RMSE of 0.165 while the MG-GNN model exhibits a very high RMSE of 1.824. Since the MG-GNN model exhibits such a high error, combining the architectures would not provide any benefits and therefore we do not report any combined results. Hence, the simple WS-GNN is able to perform above average predictions on surfactant ternary mixtures even trained exclusively on binary and pure species data. We hypothesize that the weighted summation on the WS-GNN model, ensures that the mixture fingerprint always remains in the same order of magnitude. On the other side, in the MG-GNN model, a third node is introduced in the mixture graph. After the summation pooling step, the order of magnitude of the mixture fingerprint does not remain similar to that of binary mixtures. Here, with no data on ternary mixtures during training the MG-GNN model completely fails to give any reasonable predictions. Replacing the summation with a mean pooling step led only to slight performance improvements. Thus, only the WS-GNN model could be used when dealing with ternary mixtures. We provide a parity plot with the predictions from the WS-GNN model in the SI, Figure S3.

\begin{table}[htpb]
    \caption{Error metrics of WS-GNN and MG-GNN models on the ternary mixtures data set. MAPE is in unit of \%.}
    \centering
    \begin{tabular}{l | c c c}
        Model & RMSE & MAE & MAPE \\
        \hline
         WS-GNN & \textbf{0.165} & \textbf{0.13} & \textbf{5.4} \\
         MG-GNN & 1.824 &1.687 & 74.805 \\
    \end{tabular}
    \label{tab:ternary_results}
\end{table}

\subsection{Experimental validation and commercial surfactants}\label{sec:exp_val}
\noindent
The experimentally measured mixtures comprise, binary, ternary and quaternary mixtures. The combined GNN model is utilized for binary mixtures, while the WS-GNN is used for ternary and quaternary mixtures, in accordance with the discussion above. The results are summarized in Table~\ref{tab:experiments}. In the case of commercial surfactants that are composed of two species, the combined GNN model demonstrates high accuracy, especially for the D-AB30 with an absolute error (AE) of 0.022. To better understand why the performance of the model on S-1214G, i.e., AE of 0.262, does not match the one of D-AB30, we investigate their impurities. In the case of D-AB30 only two further species exist in small concentrations. In contrast, there exist more than five further species in the S-1214G. Hence, we hypothesize the non treatment of impurities from our model as a reason for this deviation. For the ternary mixtures, we observe a MAE of 0.178 which is slightly higher than the one found in Section~\ref{sec:pred_ternaries}, namely 0.13. We note that the mixture between SDS and D-AB30 at a 0.4-0.6 composition, exhibits the highest error and when is not taken into account, the MAE reduces to 0.147. For the T-K30UP the model predictions matches the experimental value, although no quaternary mixtures were considered during training. The high purity of T-K30UP positively contributes on the model predictions. Overall, our developed GNN models can be effectively applied on commercial surfactants that are composed up to four species and help guide research and development. Accounting for the impurities is critical for further model refinement and should be addressed in a future work.

\begin{table}
    \caption{Model predictions for $log(\text{CMC})$ versus the experimental data, measured as part of this work. On parentheses the absolute CMC values in mM are provided.}
    \label{tab:experiments}
    \centering
    \begin{tabular}{l | c c c c}
        Sample & \multicolumn{3}{c}{Predicted $log(\text{CMC})$} & Measured $log(\text{CMC})$ \\
         & WS-GNN & MG-GNN & Combined & \\
        \hline
         D-AB30& 2.9 (0.79) & 2.83 (0.68) & 2.86 (0.73) & 2.84 (0.7)  \\
         S-1214G & 3.7 (5.02) & 3.59 (3.92) & 3.65 (4.47) & 3.39 (3.53) \\
         SDS (0.4)/D-AB30 (0.6) & 2.86 (0.72) & - & - & 2.75 (0.56) \\
         SDS (0.6)/D-AB30 (0.4) & 3.18 (1.51) & - & - & 2.91 (0.8) \\
         SDS (0.2)/D-AB30 (0.8) & 2.65 (0.45) & - & - & 2.55 (0.35) \\
         T-V95G & 3.68 (4.79) & - & - & 3.90 (3.59) \\
         T-K30UP & 3.44 (2.74) & - & - & 3.3 (1.98) \\
    \end{tabular}
    \label{tab:experiments}
\end{table}

\subsection{Comparison to semi-empirical model}\label{sec:semi_empirical}
\noindent

\begin{figure}[tb!]
	\captionsetup[subfloat]{font=small}
	\subfloat[Cationic / Cationic]{\includegraphics[height =5.2 cm, width = 5.4cm]{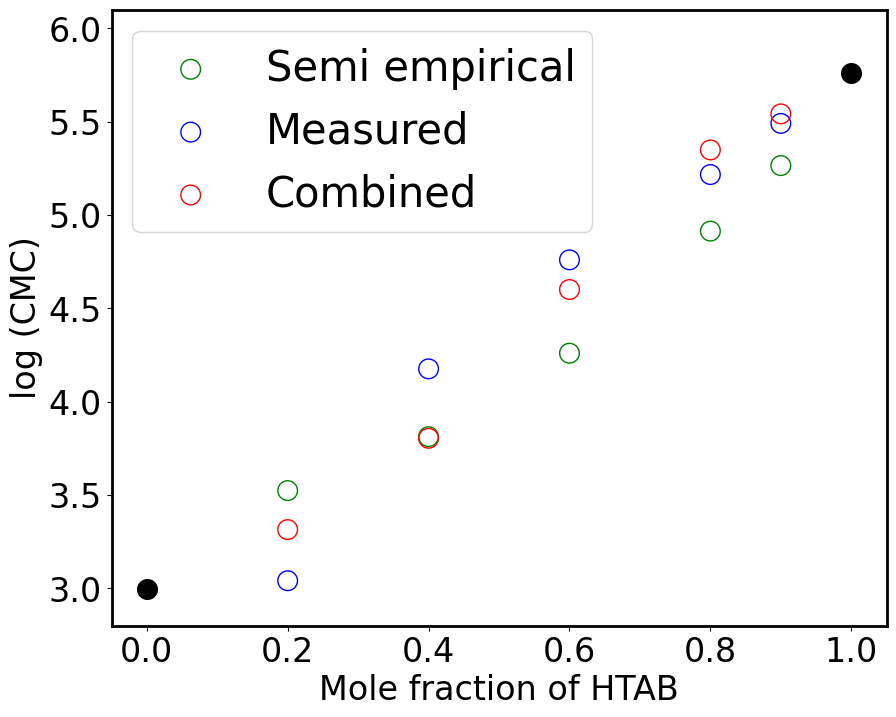}}   
	\subfloat[Zwitterionic / Cationic]{\includegraphics[height =5.2 cm, width = 5.4cm]{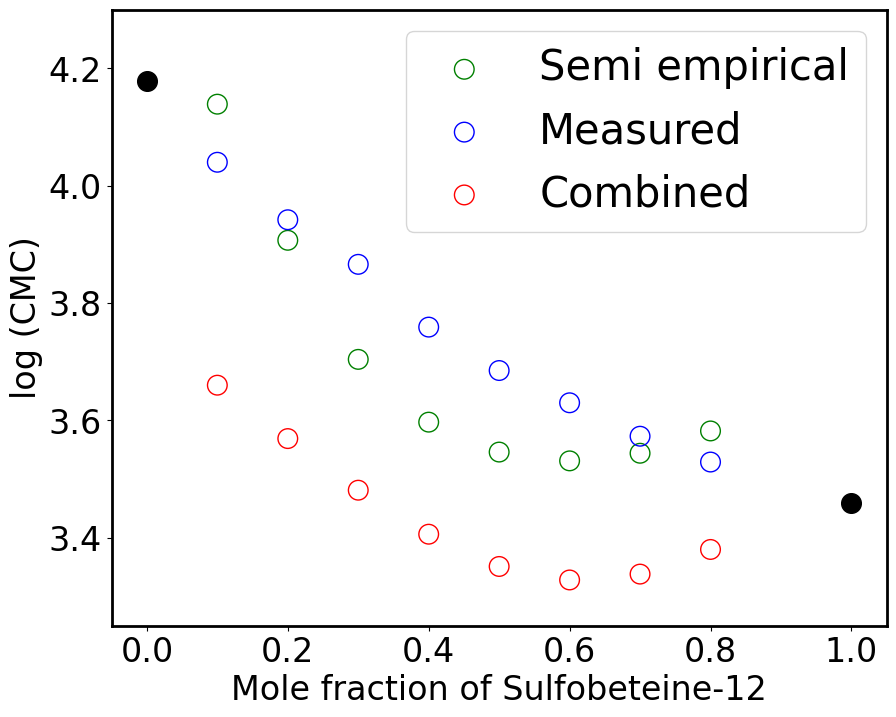}}  
	\subfloat[Cationic / Nonionic]{\includegraphics[height =5.2 cm, width = 5.4cm]{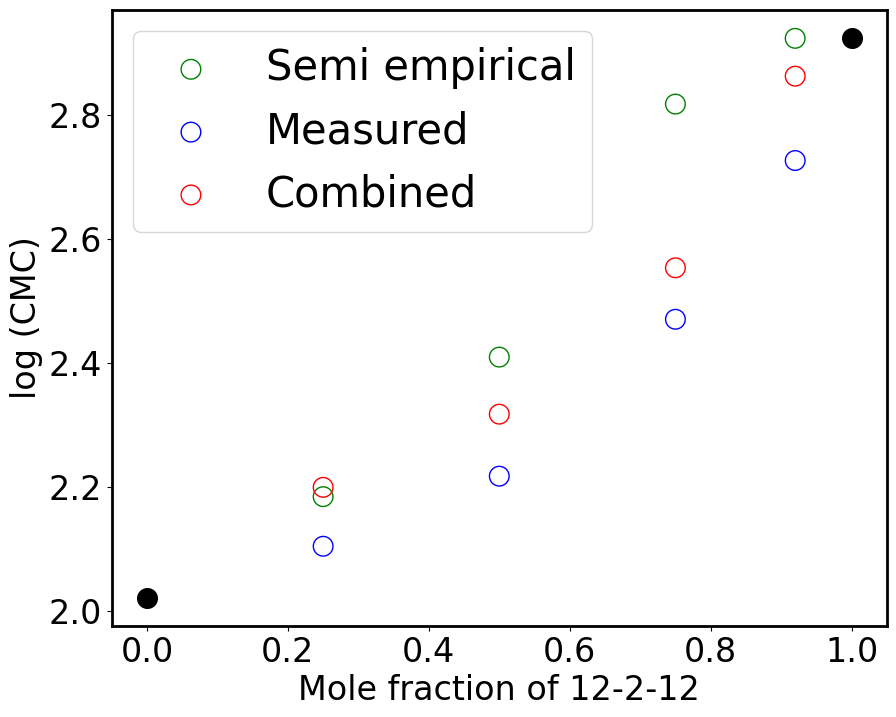}} 
	\caption{Comparison between hybrid model predictions, GNN predictions and experimental data versus on binary surfactant mixtures from the mix-comp-extra test set. Panel (a) refers to a mixture between n-hexyltrimethylammonium bromide (HTAB) and CTAB~\citep{L.LopezFontan1999}, panel (b) S-12 and DTAB~\citep{McLachlan2006} and panel (c) between 12-2-12 and C$_{12}$E$_{8}$~\citep{Alargova2001}. Black filled circles represent data points present in the training set.}
	\label{fig:hybrid_graphs}
\end{figure}

We further compare the combined GNN model to the semi-empirical model described by Eq.~\ref{eqn:activity_coef}, as we also did in a recent work of ours~\citep{Nevolianis2024}. We calculate the activity coefficients with HANNA, a hard-constraint neural network (HANNA) that ensures thermodynamic consistency that was trained on the Dortmund Data Bank, one of the largest collections of experimental activity coefficients, and was recently proposed by Specht et al.~\citep{Specht2024}. To calculate the CMCs of the two mixture species, we consider our recently developed GNN model for temperature-dependent CMCs of pure species~\citep{Brozos2024}. The results of the semi-empirical model on the logarithmic scale are given in Table~\ref{tab:hybrid_comparison} and are compared with the combined model described in Section~\ref{sec:pred_perf}. Furthermore, the predictions of both models, namely semi-empirical and combined, for three exemplary binary mixtures are graphically represented alongside the corresponding experimental measurements in Figure~\ref{fig:hybrid_graphs}.

\begin{table}[htbp]
	\caption{Comparison between the prediction accuracy of the combined GNN model versus the semi-empirical model on 2 test scenarios. MAE = mean absolute error, MAPE = mean absolute percentage error (unit \%).}
	\centering
	\begin{tabular}{c   c |  c  c}
		Model &  & mix-comp-extra  & mix-extra \\ 
		\hline
		\cline{3-4}
		Combined & \multirow{2}{*}{RMSE} & \textbf{0.313} & \textbf{0.344} \\
		Semi-empirical & & 0.568 & 0.783  \\
		\hline
		Combined & \multirow{2}{*}{MAE} & \textbf{0.196} & \textbf{0.274} \\ 
		Semi-empirical & & 0.415 & 0.621  \\
		\hline
		Combined &  \multirow{2}{*}{MAPE} &  \textbf{7.827} &  
		\textbf{10.789} \\ 
		Semi-empirical & & 15.204 & 21.995  \\
	\end{tabular}
	\label{tab:hybrid_comparison}
\end{table}

\par The results show that the combined model outperforms the semi-empirical model in both of the test scenarios, with a reduced RMSE of about half of the semi-empirical model. Investigating the three mixtures of Figure~\ref{fig:hybrid_graphs}, we observe that in all cases except for mixture (b), the combined model outperforms the semi-empirical one. Yet, the semi-empirical model accurately captures the variations of the $log(\text{CMC})$ with the mole fraction, and might be used only for a first estimate of the $CMC_\text{mix}$ given the significant higher accuracy of the combined GNN model trained on the mixture CMC data directly.

\clearpage

\section{Conclusion}\label{sec:conclusion}
\noindent 
We present a GNN-based framework for the prediction of temperature-dependent CMCs of surfactant mixtures. We collect data from literature sources for 108 binary mixtures, to which we concatenate data for pure species from our previous work. We develop and test two GNN architectures. In the first architecture, a weighted summation on the molecular fingerprints of the two mixture components is applied. In the second, a more complex mixture graph to capture inter-molecular and intramolecular interactions is proposed. We implement different test splits to uncover model capabilities and limitations for different practical applications.

\par Both GNN models exhibit very high performance in the interpolation between mixture compositions scenario, showing that GNNs are capable of accurately predicting $CMC_\text{mix}$ values at new mixture compositions. Extrapolation to new mixtures from known surfactants is also handled very well from both GNN models. Further extrapolation to mixtures where either one or both pure species of the mixture are previously unseen by the model, decreased model accuracy as expected. However, the model predictions remained accurate with some notable exceptions. Our findings indicate that no clear best-performing GNN architecture can be identified and hence we decide to combined them for more robust results, leading to overall highly accurate CMC predictions especially for the first two test scenarios. Further analysis is conducted on the predictive performance on different class combinations.

\par We further investigate if a GNN model trained on binary mixture and pure species data can be applied to ternary mixtures, without further model adjustment. We collect a small external data set of 6 ternary mixtures and test both GNN models. We find that the WS-GNN model can predict the $CMC_\text{mix}$ of ternary mixtures with high accuracy, hence demonstrating potential for further development and applications, such as consideration of the unreacted surface-active raw material. Experimental validation on 4 commercial surfactants that contain up to 4 species and ternary mixtures is conducted and very good agreement between predictions and measurements is observed. Therefore, the model can accurately screen commercial surfactants and accelerate industrial research and development in a sustainable way.

Since the CMC of surfactants, mixtures, and pure species is influenced by the pH of the solution, accounting for this factor in future work would be very interesting. This is particularly pressing in the case of zwitterionic surfactants. Furthermore, understanding the impact of different surface-active species present in the final product would allow us to either incorporate such information during model training or refine a high-fidelity data set. Extending the models to ethoxylated alkylsulfates would be highly industrial relevant, however dealing with the wide product distribution of ethoxylation remains challenging. Lastly, the lack of data availability for surfactant mixtures of higher order limits the development and/or testing of ML models for CMC prediction and should be addressed in future works. We however anticipate that the model can predict $CMC_\text{mix}$ values for mixtures with any number of surfactants, i.e., beyond quaternary systems presented in this work. The pre-requisite is that the structures of all species and the surface-active impurities are known.
\section*{Acknowledgments}

\noindent The BASF authors (C. Brozos, E. Akanny, S. Bhattacharya, and C. Kohlmann) were funded by the BASF Personal Care and Nutrition GmbH.
J. G. Rittig and A. Mitsos acknowledge funding from the Deutsche Forschungsgemeinschaft (DFG, German Research Foundation) – 466417970 – within the Priority Programme ``SPP 2331: Machine Learning in Chemical Engineering''.
Additionally, J. G. Rittig acknowledges the support of the Helmholtz School for Data Science in Life, Earth and Energy (HDS-LEE).

\section*{Data availability}
\noindent The Python scripts and the mix-comp-extra and mix-extra tests used in this work are available as open-source at our~\href{https://github.com/brozosc/Predicting-the-Temperature-Dependent-CMC-of-Surfactant-Mixtures-with-Graph-Neural-Networks}{GitHub repository}. We do not
directly provide the training set used in this work, as it remains the property of BASF, but it could be made available upon request.

\section*{Author contribution}
\noindent
\textbf{Christoforos Brozos}: Conceptualization, Methodology, Software, Data curation, Validation, Formal analysis, Experimental methodology, Writing - Original Draft, Writing - Review \& Editing, Visualization

\noindent\textbf{Jan G. Rittig}: Conceptualization, Methodology, Software, Formal analysis, Writing - Review \& Editing

\noindent\textbf{Elie Akanny}: Conceptualization, Methodology, Experimental methodology \& measurements, Writing - Review \& Editing

\noindent\textbf{Sandip Bhattacharya}: Conceptualization, Methodology, Supervision, Writing - Review \& Editing

\noindent\textbf{Christina Kohlmann}: Writing - Review \& Editing, Supervision, Funding acquisition

\noindent\textbf{Alexander Mitsos}: Writing - Review \& Editing, Supervision, Funding acquisition

\section*{Declaration of competing interest}
The authors declare the following financial interests/personal relationships which may be considered as potential competing interests: Jan Rittig and Alexander Mitsos have no competing interests. Christoforos Brozos, Elie Akanny, Sandip Bhattacharya and Christina Kohlmann are employees of BASF Personal Care and Nutrition GmbH which uses surfactants
commercially. 

  \clearpage

  \bibliographystyle{apalike}
  \renewcommand{\refname}{Bibliography}
  \bibliography{literature.bib}

\end{document}


\thispagestyle{firststyle}
	
	\begin{center}
		\begin{large}
			\textbf{\mytitle}
		\end{large} \\
		\vspace{0.1cm}
		\myauthor
	\end{center}
	
	\vspace{-0.4cm}
	
	\begin{footnotesize}
		\affil
	\end{footnotesize}
	
	\vspace{-0.3cm}

\section{Figures}
\noindent 
\begin{figure}[h]
    \centering
    \includegraphics[height =14.1cm, width = 17cm]{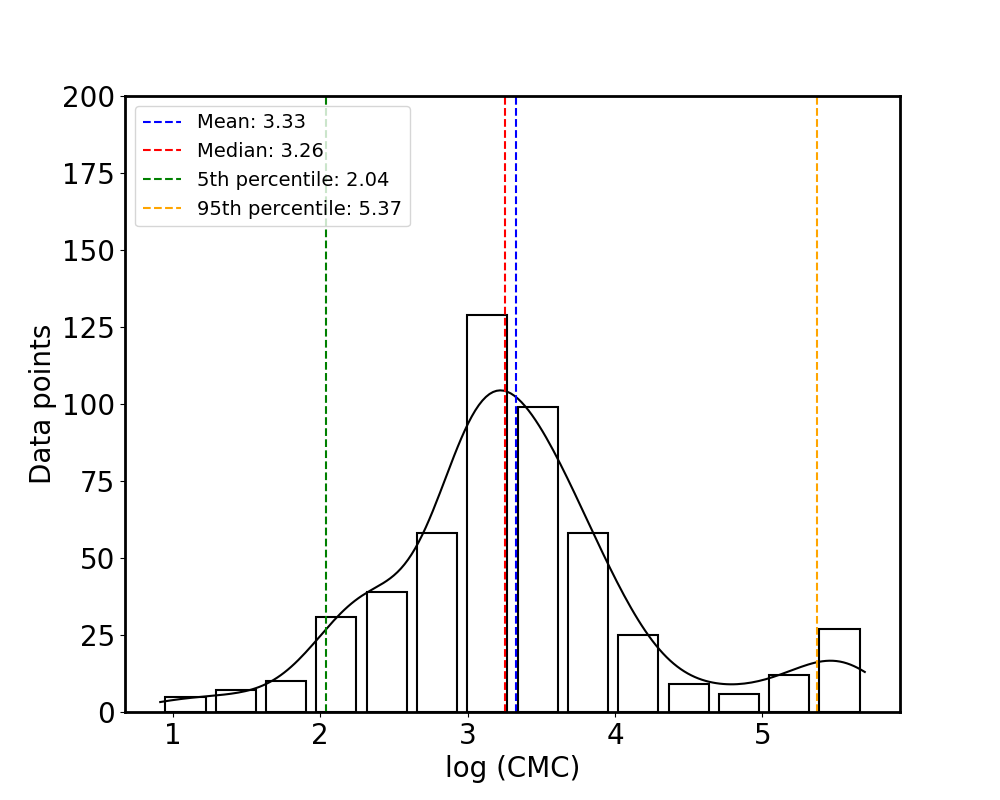}
    \caption{Statistical overview of the CMC of the binary  mixtures points. The logarithm is applied to CMC in $\mu$M.}
    \label{fig:apendix_cmc_distribution}
\end{figure}

\begin{figure}[h!]
    \subfloat[Anionic/Cationic]{\includegraphics[height =5.0 cm, width = 5.5cm]{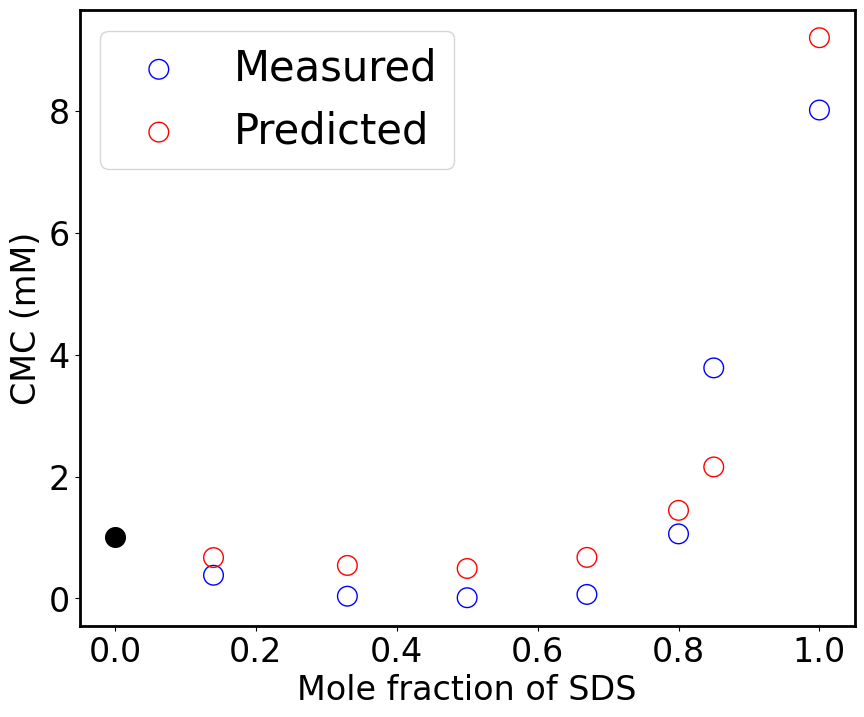}}  
    \subfloat[Anionic/Anionic]{\includegraphics[height =5.0 cm, width = 5.5cm]{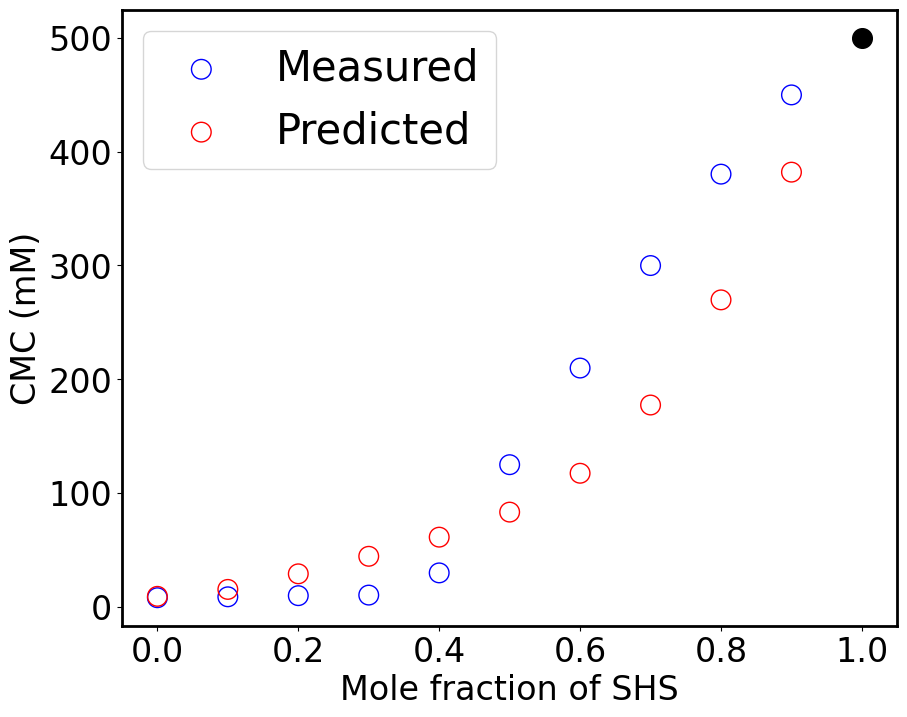}}
    \subfloat[Zwitterionic/Cationic]{\includegraphics[height =5.0 cm, width = 5.5cm]{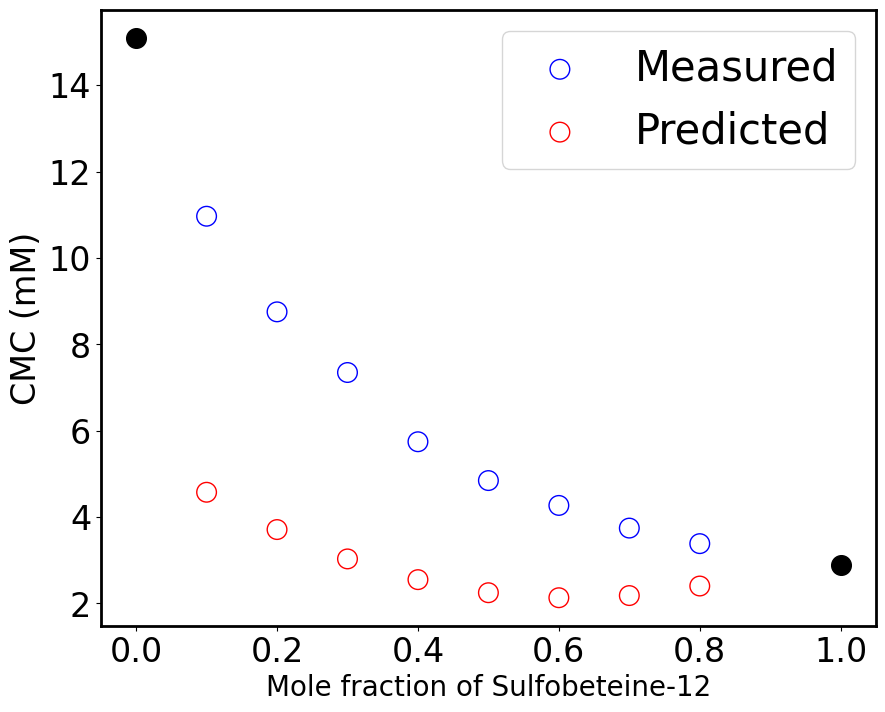}}
    \caption{GNN predictions versus experimental data on different surfactant mixtures present in two test scenarios. Panel (a) refers to a mixture between SDS and CPC~\citep{Maiti2010}, (b) refers to a mixture between SHS and SDS~\citep{LopezFontan2000} and (c) to a mixture between S-12 and DTAB~\citep{McLachlan2006}. Black filled circles represent data points present in the training set.}
    \label{fig:nonionic_nonionic}
\end{figure}

\begin{figure}
    \centering
    \includegraphics[height =7.5 cm, width = 8.5cm]{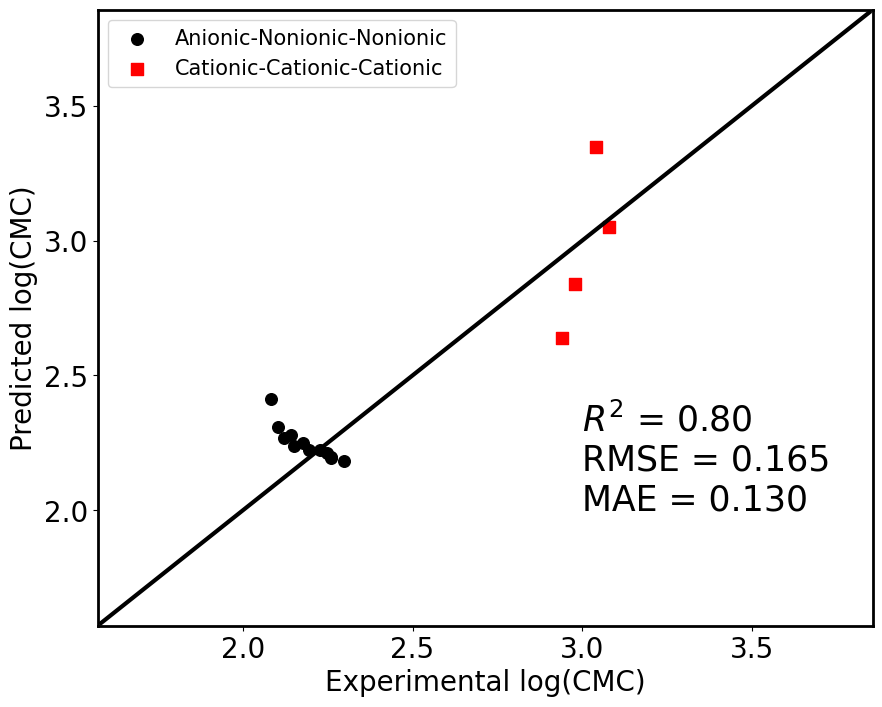}  
    \caption{Parity plot of the WS-GNNs on the ternary mixture data set. The data points are highlighted with different colors and markers based on the classes of the three mixture components. The logarithm is applied to CMC in $\mu$M (base 10)}
    \label{fig:parity_plot_ternary}
\end{figure}

\clearpage
\section{Tables}

\begin{table}[h]
    \caption{CMC values of dodecyltrimethylammonium bromide collected from literature sources at 30~$^\circ$C. The CMC values are averaged out. Note, that the fifth CMC entry on the above table was measured at 29.85~$^\circ$C~\citep{Ray2005}. We rounded such temperature entries to full numbers, e.g., in this case to 30~$^\circ$C.}
    \centering
    \begin{tabular}{c|c}
       \textbf{Source}  & \textbf{CMC (mM)}  \\
       \hline
         ~\citep{Chauhan2014} & 15.5 \\
         ~\citep{YogeshKadam2009} & 15\\
         ~\citep{L.LopezFontan1999} & 15.6\\
         ~\citep{Hierrezuelo2006} & 15\\
         ~\citep{Ray2005} & 15.7 \\
        \hline
         \textbf{Average} &\textbf{15.36}
    \end{tabular}
    \label{tab:appendix_cmc_values}
\end{table}

\begin{table}
    \caption{Number of mixtures of the different surfactant combinations in the full mixture data set and in each of the four test sets. The number of data points is given in parentheses. Note that for mix-compo-extra and mix-surf-extra test sets, the aggregated number of data points is given.}
    \centering
    \begin{tabular}{c| c | c c c c}
        Combination  & Full mixture data set  &  Comp-inter & Mix-comp-extra & Mix-surf-extra & Mix-extra  \\
        \hline
        An.-Non. & 11 & 9 & 1 & 5 & 2\\
        An.-Cat. & 13 & 8 & 1 & 1 & 0\\
        An.-An. & 8 & 8 & 0& 4 & 0\\
        An.-Zwitt. & 6 & 6& 1& 5 & 0\\
        Cat.-Non. & 24 & 23 & 4 & 9 & 0 \\
        Cat.-Cat. & 35 & 32 & 8& 5& 2\\
        Cat.-Zwitt. & 5 & 5 & 2 & 0 &0\\
        Non.-Non. & 5 & 4 & 2 & 1 & 3\\
        Non.-Zwitt. & 1 & 1 & 0 & 1 & 0 \\
        \textbf{Total} & \textbf{108} (515) & \textbf{96} (96) & \textbf{19} (104) & \textbf{25} (170) & \textbf{7} (38)
    \end{tabular}
    \label{tab:number_of_mixtures}
\end{table}

\begin{table}
    \caption{Hyperparameters of the GNN model investigated through a grid search. The hyperparameter dimensions refers to the size of the molecular fingerprint and the size of the MLP.}
    \begin{center}
        \begin{tabular}{c|c |c}
             \textbf{Hyperparameter}& \textbf{Range} & \textbf{Optimized architecture}\\
             \hline
             Graph convolutional layers & 1 & \textbf{1}\\
             Graph convolutional type & GINEConv & \textbf{GINEConv}\\
             Learnable parameter $\in$ & (True, False) & \textbf{False} \\
             Initial learning rate & (0.001, 0.005, 0.01) & \textbf{0.001} \\
             Batch size & (32, 64, 128) & \textbf{32}\\
             Dimensions & (64, 128) & \textbf{128}\\
             Number of MLP layers & 3 & \textbf{3} \\
             Activation function & ReLU & \textbf{ReLU} \\
             Maximum epochs & 200 & \textbf{200} \\
             Early stopping patience & 30 & \textbf{30} \\
             Learning rate decay & 0.8 & \textbf{0.8} \\
             Patience & 3 & \textbf{3} \\
             Optimizer & Adam & \textbf{Adam}\\
        \end{tabular}
        \label{tab:appendix_model_hyperparameters}
    \end{center}
\end{table}

    \clearpage
  \bibliographystyle{apalike}
  \renewcommand{\refname}{Bibliography}
  \bibliography{literature.bib}